\documentclass{elsart}
\usepackage{epsfig}
\usepackage{latexsym}
\usepackage{amsmath}
\usepackage{amssymb}
\newcommand{\keyw}[1]{\hbox{\bf #1}}
\newcommand{\floor}[1]{{\lfloor{#1}\rfloor}}
\newcommand{\FT}{{Fault\kern1pt-\kern-2pt Tolerant}}
\def\nmr{\hbox{$N$\kern-1pt MR}}
\begin{document}
\begin{frontmatter}
\title{The Algorithm of Pipelined Gossiping}
%
\author{Vincenzo De~Florio and Chris Blondia}
%
\address{University of Antwerp\\
 Department of Mathematics and Computer Science\\
 Performance Analysis of Telecommunication Systems group\\
 Middelheimlaan 1, 2020 Antwerp, Belgium, \emph{and}\\
 Interdisciplinary institute for BroadBand Technology\\
 Crommenlaan 8, 9050 Ghent-Ledeberg, Belgium.}
\end{frontmatter}

\noindent
\centerline{\bf Abstract}

\noindent
A family of gossiping algorithms depending on a parameter permutation
is introduced, formalized, and discussed. Several of its members
are analyzed and their asymptotic behaviour is revealed, including
a member whose model and performance closely follows the one of
hardware pipelined processors. This similarity is exposed. An optimizing
algorithm is finally proposed and discussed as 
a general strategy to increase the
performance of the base algorithms.
%

\section{Introduction}
A number of distributed applications like, e.g.,
distributed consensus~\cite{LaSP82}, or those based on the concept
of restoring organs~\cite{DeDL00c,John89a} ($N$-modular redundancy systems
with $N$-replicated voters---for instance, the distributed voting
tool described in~\cite{DeDL98e}),
require a base service called gossiping~\cite{Goss,BGRV98,Gon03}.

Informally speaking, gossiping is a communication procedure such that
every member of a set has to communicate a private value to all
the other members. Gossiping is clearly an expansive service,
as it requires a large amount of communication. Implementations
of this service can have a great impact on the throughput
of their client applications and perform very differently
depending on the number of members in the set.
This work describes a family of gossiping algorithms that
depend on a combinatorial parameter.
Three cases are then analyzed under the hypotheses of discrete
time, of constant time for performing a \verb"send" or
\verb"receive", and of a crossbar communication system.
It is shown how, depending on the pattern of the parameter,
gossiping can use from $\hbox{O}(N^2)$ to $\hbox{O}(N)$ time,
$N$ being the number of communicating members.
The last and best-performing case, whose activity follows the
execution pattern of pipelined hardware processors, is shown
to exhibit an efficiency constant with respect to $N$.
This translates in unlimited scalability of the corresponding
gossiping service. When performing multiple consecutive gossiping
sessions, the throughput of the system can reach the value
of $t/2$, $t$ being the time for sending one value from
one member to another, or a full gossiping is completed
every two basic communication steps.

The structure of the paper follows:
first, in Sect.~\ref{model}, a formal model for the family of algorithms is provided.
The following three sections (Sect.~\ref{zerocase}, Sect.~\ref{randcase}, and Sect.~\ref{tao-pb})
introduce, analyze, and discuss 
three members of the family, showing in particular that one of them,
whose behaviour resembles the one of pipelined hardware microprocessors,
uses $\hbox{O}(N)$ time, $N$ being the number of employed nodes.
An optimizing algorithm is then introduced in Sect.~\ref{opt}.
Section~\ref{vf} describes two applications of our algorithms.
Finally Sect.~\ref{end} summarizes our contributions 
and draws a number of conclusions.

\section{A Formal Model}\label{model}
\begin{defn}[system]
Let $N>0$. $N+1$ processors are interconnected via some communication
means that allows them to communicate with each other (for instance, by means
of full-duplex point-to-point communication lines).
Communication is synchronous and blocking.
Processors are uniquely identified by integer labels in $\{0,\dots,N\}$;
they will be globally referred to, together with the communication
means, as ``the system''.
\end{defn}

\begin{defn}[problem]\label{problem}
The processors own some local data
they need to share (for instance, to execute a voting algorithm~\emph{\cite{John89a}}).
In order to share their local data, each processor needs to
broadcast its own data to all others, via multiple sending operations,
and to receive the $N$ data items
owned by its fellows. This must be done as soon as possible.
We assume a discrete time model---events occur at discrete time
steps, one event at a time per processor. 
This is a special class of the general family
of problems of information dissemination
known as \emph{gossiping\/}~\emph{\cite{Goss,BGRV98,Gon03}}. We will refer to this class
as ``the problem''.
\end{defn}

\begin{defn}[time step]
\label{timestep}
We assume the time to send a message and that to receive a message
is constant. We call this amount of time a ``\emph{time step}''.
\end{defn}

\begin{defn}[actions]
On a given time step $t$, processor $i$ may be:
  \begin{enumerate}
    \item sending a message to processor $j, j\neq i$; 
    this is represented in form of relation as $i\, S^t j$;
    \item receiving a message from processor $j, j\neq i$; 
    this is represented as $i\, R^t j$;
    \item blocked, waiting for messages to be \emph{received\/} from any processor; 
    where both the identities
    of the involved processors and $t$ can be omitted without
    ambiguity, symbol ``$-$'' will be used to represent this case;
    \item blocked, waiting for a message to be \emph{sent\/} i.e., for a designee
    to enter the receiving state;
    under the same assumptions of case (3), symbol ``$\curvearrowright$''
    will be used.
  \end{enumerate}
The above cases are referred to as ``the actions'' of a time step.
\end{defn}

\begin{defn}[slot, used slot, wasted slot]
A slot is a temporal ``window'' one time step long, related to a processor.
On each given time step there are $N+1$ available slots within the system.
Within that time step, a processor may \emph{use\/} that slot 
(if it sends or receives a message during that slot), or it may 
\emph{waste\/} it (if it is in one of the remaining two cases). In other words:

\noindent
Processor $i$ \emph{makes use\/} of slot $t$ 
(represented by predicate $U(t,i)$)
if and only if
\[
  U(t,i) = \textrm{``}\exists \, j \, (i\, S^t j \vee i\, R^t j)\textrm{''}
\]
\noindent
is true; on the contrary, processor $i$ is said to \emph{waste\/} slot $t$ if{}f $\neg U(t,i)$.

\noindent
The following notation,
\begin{equation}\label{u}
\delta_{i,t} =
         \begin{cases}
         1&\text{if $U(t,i)$ is true},\\
         0&\text{otherwise},
         \end{cases}
\end{equation}
will be used to count used slots.
\end{defn}

\begin{defn}[states $W\!R, W\!S, S, R$]
Let us define four state templates for a finite state automaton 
(FSA) to be described later on.

\begin{description}
\item[$W\!R$ state.]
A processor is in state $W\!R_j$ if it is waiting for the arrival of a message
from processor $j$. Where the subscript is not important it will be omitted.
Once there, a processor stays in state $W\!R$ for \emph{zero\/} (if it can start
receiving immediately) or more time steps, corresponding to the same number
of actions ``wait for a message to come.''

\item[$S$ state.]
A processor is in state $S_j$ when it is sending a message to addressee processor $j$.
Note that by the above assumptions and definitions this transition lasts exactly one time step.
To each transition to the $S$ state there corresponds exactly one ``send'' action.

\item[$W\!S$ state.]
A processor which is willing to send a message to processor $j$ is said to be in
state $W\!S_j$. Where the subscript is not important it will be omitted.
The permanence of a processor in state $W\!S$ implies \emph{zero\/} (if
the processor can send immediately) or more occurrences in a row of
the ``wait for sending'' action.

\item[$R$ state.]
A processor which is receiving a message from processor $j$ is said to be in
state $R_j$. By the above definitions, this state transition also lasts 
one time step.
\end{description}
\end{defn}

Let
${\mathcal P}_1,\dots,{\mathcal P}_N$ represent a permutation of the 
$N$ integers $0,\dots,i-1,i+1,\dots,N$. 
Then the above state templates can be used to compose $N+1$
finite state automata making use of the following
algorithm ($i\in\{0,\dots,N\}$):

\begin{alg}{\emph{: Compose the FSA which solves the problem of Def.~\ref{problem} for processor $i$}}
\begin{tabbing}
xx \= xx \= xx \= xx \= xx \= xx \= xx \= xx \= xx \= xx \= xx \= xx \= xxxxx \= xx \kill
         \> \> \emph{Input:\/} $A \equiv (i, N, \mathcal P)$\\
         \> \> \emph{Output:\/} $\mathrm{FSA}(A)$\\
{\bf  1} \> \> \keyw{begin}\\
{\bf  2} \> \> \> $\mathrm{FSA}(A) := \mathrm{START}$ \>\>\>\>\>\>\>\>\>\> \{ emit the initial state \} \\
{\bf  3} \> \> \> \keyw{for} $j := 0$ \keyw{to} $i-1$ \keyw{do} \\
         \> \> \> \> \{ operator ``$\rightarrow$'' pushes a state on top of a FSA \} \\
{\bf  4} \> \> \> \> $\mathrm{FSA}(A) := \mathrm{FSA}(A) \rightarrow W\!R$ \\
{\bf  5} \> \> \> \> $\mathrm{FSA}(A) := \mathrm{FSA}(A) \rightarrow R$ \\
{\bf  6} \> \> \> \keyw{enddo}\\
{\bf  7} \> \> \> \keyw{for} $j := 1$ \keyw{to} $N$ \keyw{do} \\
{\bf  8} \> \> \> \> $\mathrm{FSA}(A) := \mathrm{FSA}(A) \rightarrow W\!S_{{\mathcal P}_j}$\\
{\bf  9} \> \> \> \> $\mathrm{FSA}(A) := \mathrm{FSA}(A) \rightarrow S_{{\mathcal P}_j}$\\
{\bf 10} \> \> \> \keyw{enddo}\\
{\bf 11} \> \> \> \keyw{for} $j := i+1$ \keyw{to} $N$ \keyw{do}\\
{\bf 12} \> \> \> \> $\mathrm{FSA}(A) := \mathrm{FSA}(A) \rightarrow W\!R$\\
{\bf 13} \> \> \> \> $\mathrm{FSA}(A) := \mathrm{FSA}(A) \rightarrow R$\\
{\bf 14} \> \> \> \keyw{enddo}\\
{\bf 15} \> \> \> $\mathrm{FSA}(A) := \mathrm{FSA}(A) \rightarrow \mathrm{STOP}$ \>\>\>\>\>\>\>\>\>\> \{ emit the final  state \} \\
{\bf 16} \> \> \keyw{end}.
\end{tabbing}
\label{alg1}
\end{alg}

\begin{figure}
\centerline{\psfig{figure=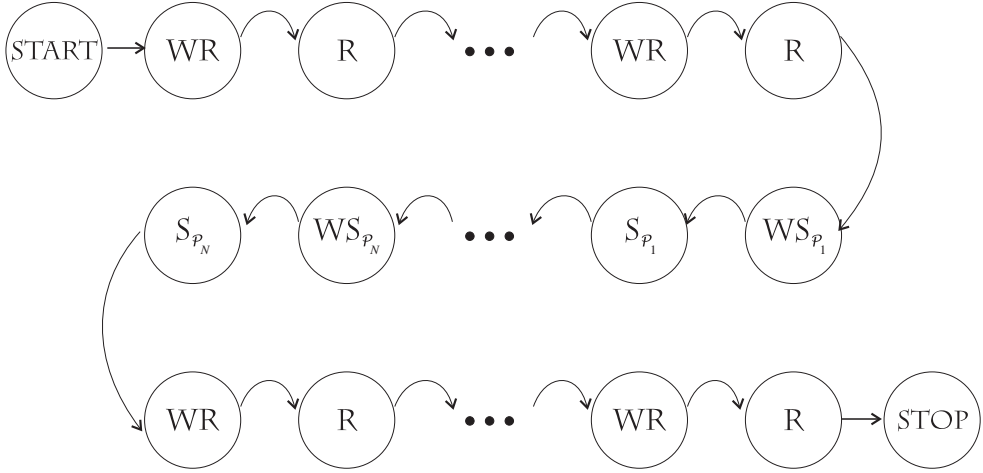,width=13cm}}
\caption{The state diagram of the FSA run by processor $i$. The first
row consists of $i$ couples \( (W\!R, R) \).
$({\mathcal P}_1,\dots,{\mathcal P}_N)$ represents a permutation of the $N$ integers
$0,\dots,i-1,i+1,\dots,N$.
The last row contains $N-i$ couples \( (W\!R, R) \).
}\label{fsa}
\end{figure}

Figure~\ref{fsa} for instance shows the state diagram of the FSA
to be executed by processor $i$.
The first row represents the condition that has to be reached before 
processor $i$ is allowed to begin its broadcast: a series 
of $i$ couples \( (W\!R, R) \).

Once processor $i$ has successfully received $i$ messages, it gains 
the right to broadcast, which it does according to the rule expressed in 
the second row of Fig.~\ref{fsa}: it orderly sends its message to
its fellows, the $j$-th message being sent to processor ${\mathcal P}_j$.

The third row of Fig.~\ref{fsa} represents the reception of the remaining $N-i$ messages, 
coded as $N-i$ couples like those in the first row.

We experimentally observed that, regardless the value of $\mathcal P$, such FSA's
represent a distributed algorithm which solves the problem of
Definition~\ref{problem} without deadlocks.
As intuition may suggest, the choice of which permutation to use has
indeed a deep impact
on the overall performance of the algorithm, together
with the physical characteristics of the communication 
line\footnote{For instance, in case of a bus, 
	      an ALOHA system (see e.g., \cite{Tan96}), or
	      other shared medium systems,
              a number of used slots greater than 2 implies a collision 
              i.e., a penalty that wastes the current slot; using
	      transputers~\cite{Gra90}, each of which has
	      four independent communication channels,
              used slots cannot be more than 8; while in a 
              fully interconnected end-to-end system, that figure
              can grow up to its maximum value, $2\floor{(N+1)/2}$, 
	      without any problem.}.
Reporting on this impact is one of the aims of this paper. 

To this end, let us furthermore define:

\begin{defn}[run]
The collection of slots needed to fully execute the above algorithm on a 
given system, together with the value of the corresponding actions.
\end{defn}

\begin{defn}[average slot utilization]
The average number of slots used during a time step. It represents
the average degree of parallelism exploited in the system.
It will be indicated as $\mu_N$, or simply as $\mu$. It varies between 0
and $N+1$.
\end{defn}

\begin{defn}[efficiency]
The percentage of used slots over the total number of slots 
available during a run. $\varepsilon_N$, or more simply $\varepsilon$, will
be used to represent efficiency.
\end{defn}

\begin{defn}[length]
The number of time steps in a run. It represents a measure
of the time needed by the distributed algorithm to complete.
$\lambda_N$, or more simply $\lambda$, will be used for lengths.
\end{defn}

\begin{defn}[number of slots]
$\sigma(N) = (N+1)\lambda_N$ represents the number of slots 
available within a run of $N+1$ processors.\label{sigma}
\end{defn}

\begin{defn}[number of used slots]
\label{nut}
For each run and each time step $t$,
\[ \nu_t = \sum_{i=0}^N \delta_{i,t} \]
represents the number of slots that have been used during $t$.
\end{defn}

\begin{defn}[utilization string]
\label{us}
The $\lambda$-tuple
\[ \vec{\nu} = [ \nu_1, \nu_2, \dots, \nu_\lambda ], \]
orderly representing the number of used slots for
each time step, is called utilization string.
\end{defn}

In the next Sections, we introduce and discuss three cases
of $\mathcal P$. We will show how varying the structure of $\mathcal P$
may develop extremely different values of $\mu$, $\varepsilon$, and $\lambda$.
This fact, coupled with physical constraints pertaining the communication line
and with the number of available independent channels,
determines the overall performance of this algorithm.

In the following we assume the availability of
a fully connected (crossbar) interconnection~\cite{PaHe96}
that allows any processor to communicate with any other processor
in one time step.

\section{First Case: Identity Permutation}\label{zerocase}
As a first case, let us assume that the structure of $\mathcal P$ 
be fixed. For instance,
let $\mathcal P$ be equal to the identity permutation:
\begin{equation}\label{zeroperm}
\binom{0,\dots,i-1,i+1,\dots,N}{0,\dots,i-1,i+1,\dots,N},
\end{equation}
i.e., in cycle notation~\cite{Knu73a},
\(
(0)\dots(i-1)(i+1)\dots(N).
\)

This means that, once processor $i$ gains the right to broadcast,
it will first try to send its message to processor 0 
(possibly having to wait for it to become available to receive that message),
then it will do the same with processor 1, and so
forth up to $N$, obviously skipping itself. This is effectively
represented in Table~\ref{run4} for $N=4$. Let us call this a run-table.

\begin{table}
\begin{center}
\begin{tabular}{l|c@{\hspace{1pt}}c@{\hspace{1pt}}c@{\hspace{1pt}}c@{\hspace{1pt}}c@{\hspace{1pt}}c@{\hspace{1pt}}c@{\hspace{1pt}}c@{\hspace{1pt}}c@{\hspace{1pt}}c@{\hspace{1pt}}c@{\hspace{1pt}}c@{\hspace{1pt}}c@{\hspace{1pt}}c@{\hspace{1pt}}c@{\hspace{1pt}}c@{\hspace{1pt}}c@{\hspace{1pt}}c@{\hspace{1pt}}}
$\stackrel{\hbox{\sffamily\scriptsize id}}{\downarrow}$ 
$\stackrel{\hbox{\sffamily\scriptsize step}}{\rightarrow}$
&\tiny1&\tiny2&\tiny3&\tiny4&\tiny5&\tiny6&\tiny7&\tiny8&\tiny9&\tiny10&\tiny11&\tiny12&\tiny13&\tiny14&\tiny15&\tiny16&\tiny17&\tiny18\\ \hline
\sf 0&$S_{1}$&$S_{2}$&$S_{3}$&$S_{4}$&$R_{1}$&$-$&$R_{2}$&$-$&$-$&$-$&$R_{3}$&$-$&$-$&$-$&$R_{4}$&$-$&$-$&$-$\\
\sf 1&$R_{0}$&$\curvearrowright$&$\curvearrowright$&$\curvearrowright$&$S_{0}$&$S_{2}$&$S_{3}$&$S_{4}$&$R_{2}$&$-$&$-$&$R_{3}$&$-$&$-$&$-$&$R_{4}$&$-$&$-$\\
\sf 2&$-$&$R_{0}$&$-$&$-$&$-$&$R_{1}$&$S_{0}$&$\curvearrowright$&$S_{1}$&$S_{3}$&$S_{4}$&$-$&$R_{3}$&$-$&$-$&$-$&$R_{4}$&$-$\\
\sf 3&$-$&$-$&$R_{0}$&$-$&$-$&$-$&$R_{1}$&$-$&$-$&$R_{2}$&$S_{0}$&$S_{1}$&$S_{2}$&$S_{4}$&$-$&$-$&$-$&$R_{4}$\\
\sf 4&$-$&$-$&$-$&$R_{0}$&$-$&$-$&$-$&$R_{1}$&$-$&$-$&$R_{2}$&$-$&$-$&$R_{3}$&$S_{0}$&$S_{1}$&$S_{2}$&$S_{3}$\\
\hline
$\stackrel{\hbox{\sffamily\scriptsize $\vec\nu$}}{\rightarrow}$&2&2&2&2&2&2&4&2&2&2&4&2&2&2&2&2&2&2
\end{tabular}
\end{center}
\caption{A run ($N=4$), with $\mathcal P$ equal to the identity permutation.
The {\sf step} row represents time steps. {\sf Id}'s identify 
processors. $\vec\nu$ is the utilization string (see Def.~\ref{us}.)
In this case
$\mu$, or the average utilization is 2.22 slots out of 5, 
with an efficiency $\varepsilon=44.44\%$ and a length $\lambda=18$.
Note that, if the slot is used, then entry $(i,t)={\mathcal R}_j$ of this matrix
represents relation $i\, {\mathcal R}^t j$.}\label{run4}
\end{table}

It is possible to characterize precisely the duration of
the algorithm adopting this permutation: 
\begin{prop}\label{prop1}
$\lambda_N = \frac34N^2 + \frac54N + \frac12 \floor{N/2}.$
\end{prop}
\begin{pf*}{PROOF (by induction)}
Let us consider run-table $N+1$. Let us strip off its last row; 
then wipe out 
the $\floor{(N+1)/2}-1$ leftmost columns which contain element $S_{N+1}$. 
Let us also cut out the whole right part of the table starting
at the column containing the last occurrence of $S_{N+1}$. Finally, let us rename all
the remaining $S_{N+1}$'s as ``$-$''.

Our first goal is showing that what remains is run-table $N$.
To this end, let us first point out how
the only actions that affect the content of other cells in a run-table
are the $S$ actions. Their range of action is given by their subscript:
an $S_{N+1}$ for instance only affects an entry in row $N+1$.

Now consider what happens when processor $i-1$ sends its message to 
processor $i$ and this latter gains the right to broadcast as well:
at this point, processor $i$ starts sending to processors in the
range $\{0,\dots,i-2\}$ i.e., those ``above''; as soon as it
tries to reach processor $i-1$, in the case this latter has not
finished its own broadcast, $i$ enters state $W\!S$ and blocks.

This means that:
\begin{enumerate}
\item processors ``below'' processor $i$ will not be allowed
to start their broadcast, and
\item for processor $i$ and those ``above'', $\mu$, or the degree of parallelism,
is always equal to 2 or 4---no other value is possible.
This is shown for instance in Table~\ref{run4}, row ``$\vec\nu\,$''.
\end{enumerate}

\begin{figure}[h]
\centerline{\psfig{figure=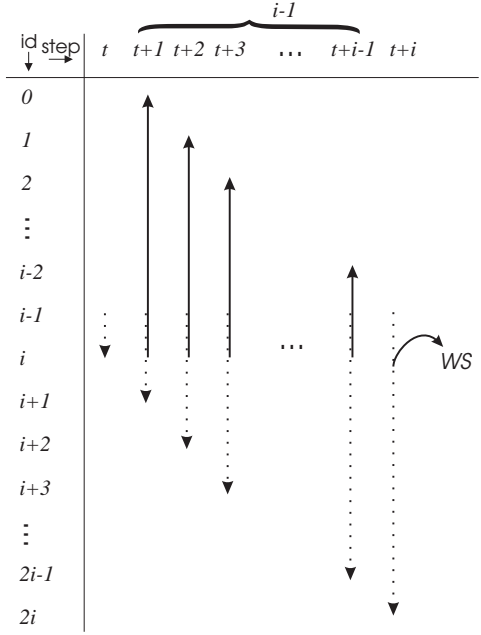,height=8.4cm}}
\caption{Processor $i-1$ blocks processor $i$ only if $2i-1<N$. 
A transmission i.e.,  two used slots, is represented by an arrow.
In dotted arrows the sender is processor $i-1$, for normal arrows
it is processor $i$.
Note the cluster of $i-1$ columns with two concurrent transmissions
(adding up to 4 used slots) in each of them.}
\label{f-prop1}
\end{figure}

As depicted in Fig.~\ref{f-prop1}, processor $i$ gets blocked
only if it tries to send to processor $i-1$ while this latter
is still broadcasting, which happens
when $i<\frac{N+1}{2}$---this condition is true for any
processor $j\in\{1,\dots, \floor{\frac{N+1}{2}}-1\}$. Note how
a ``cluster'' appears, consisting of $j-1$ columns with 4 used slots inside
(Table~\ref{runtable7} can be used to verify the above when $N$ is 7.)
Removing the first $\floor{\frac{N+1}{2}}$ occurrences
of $S_{N+1}$ (from row 0 to row $\floor{\frac{N+1}{2}}-1$)
therefore simply shortens of one time step the stay
of each processor in their current waiting states.
All remnant columns containing that element cannot be removed---these
occurrences simply vanish by substituting them with a ``$-$'' action.

\begin{table}
\begin{tabular}{l|c@{\hspace{1pt}}c@{\hspace{1pt}}c@{\hspace{1pt}}c@{\hspace{1pt}}c@{\hspace{1pt}}c@{\hspace{1pt}}c@{\hspace{1pt}}c@{\hspace{1pt}}c@{\hspace{1pt}}c@{\hspace{1pt}}c@{\hspace{1pt}}c@{\hspace{1pt}}c@{\hspace{1pt}}c@{\hspace{1pt}}c@{\hspace{1pt}}c@{\hspace{1pt}}c@{\hspace{1pt}}c@{\hspace{1pt}}c@{\hspace{1pt}}c@{\hspace{1pt}}c@{\hspace{1pt}}c@{\hspace{1pt}}c@{\hspace{1pt}}c@{\hspace{1pt}}}
$\stackrel{\hbox{\sffamily\scriptsize id}}{\downarrow}$ $\stackrel{\hbox{\sffamily\scriptsize step}}{\rightarrow}$&\tiny1&\tiny2&\tiny3&\tiny4&\tiny5&\tiny6&\tiny7&\tiny8&\tiny9&\tiny10&\tiny11&\tiny12&\tiny13&\tiny14&\tiny15&\tiny16&\tiny17&\tiny18&\tiny
19&\tiny20&\tiny21&\tiny22&\tiny23&\tiny24\\ \hline
\sf 0&$S_{1}$&$S_{2}$&$S_{3}$&$S_{4}$&$S_{5}$&$S_{6}$&$S_{7}$&$R_{1}$&$-$&$R_{2}$&$-$&$-$&$-$&$-$&$-$&$-$&$R_{3}$&$-$&$-$&$-$&$-$&$-$&$R_{4}$&$-$\\
\sf 1&$R_{0}$&$\curvearrowright$&$\curvearrowright$&$\curvearrowright$&$\curvearrowright$&$\curvearrowright$&$\curvearrowright$&$S_{0}$&$S_{2}$&$S_{3}$&$S_{4}$&$S_{5}$&$S_{6}$&$S_{7}$&$R_{2}$&$-$&$-$&$R_{3}$&$-$&$-$&$-$&$-$&$-$&$R_{4}$\\
\sf 2&$-$&$R_{0}$&$-$&$-$&$-$&$-$&$-$&$-$&$R_{1}$&$S_{0}$&$\curvearrowright$&$\curvearrowright$&$\curvearrowright$&$\curvearrowright$&$S_{1}$&$S_{3}$&$S_{4}$&$S_{5}$&$S_{6}$&$S_{7}$&$R_{3}$&$-$&$-$&$-$\\
\sf 3&$-$&$-$&$R_{0}$&$-$&$-$&$-$&$-$&$-$&$-$&$R_{1}$&$-$&$-$&$-$&$-$&$-$&$R_{2}$&$S_{0}$&$S_{1}$&$\curvearrowright$&$\curvearrowright$&$S_{2}$&$S_{4}$&$S_{5}$&$S_{6}$\\
\sf 4&$-$&$-$&$-$&$R_{0}$&$-$&$-$&$-$&$-$&$-$&$-$&$R_{1}$&$-$&$-$&$-$&$-$&$-$&$R_{2}$&$-$&$-$&$-$&$-$&$R_{3}$&$S_{0}$&$S_{1}$\\
\sf 5&$-$&$-$&$-$&$-$&$R_{0}$&$-$&$-$&$-$&$-$&$-$&$-$&$R_{1}$&$-$&$-$&$-$&$-$&$-$&$R_{2}$&$-$&$-$&$-$&$-$&$R_{3}$&$-$\\
\sf 6&$-$&$-$&$-$&$-$&$-$&$R_{0}$&$-$&$-$&$-$&$-$&$-$&$-$&$R_{1}$&$-$&$-$&$-$&$-$&$-$&$R_{2}$&$-$&$-$&$-$&$-$&$R_{3}$\\
\sf 7&$-$&$-$&$-$&$-$&$-$&$-$&$R_{0}$&$-$&$-$&$-$&$-$&$-$&$-$&$R_{1}$&$-$&$-$&$-$&$-$&$-$&$R_{2}$&$-$&$-$&$-$&$-$\\
\hline
$\stackrel{\hbox{\sffamily\scriptsize $\vec\nu$}}{\rightarrow}$&2&2&2&2&2&2&2&2&2&4&2&2&2&2&2&2&4&4&2&2&2&2&4&4
\end{tabular}
\begin{tabular}{l|c@{\hspace{1pt}}c@{\hspace{1pt}}c@{\hspace{1pt}}c@{\hspace{1pt}}c@{\hspace{1pt}}c@{\hspace{1pt}}c@{\hspace{1pt}}c@{\hspace{1pt}}c@{\hspace{1pt}}c@{\hspace{1pt}}c@{\hspace{1pt}}c@{\hspace{1pt}}c@{\hspace{1pt}}c@{\hspace{1pt}}c@{\hspace{1pt}}c@{\hspace{1pt}}c@{\hspace{1pt}}c@{\hspace{1pt}}c@{\hspace{1pt}}c@{\hspace{1pt}}c@{\hspace{1pt}}c@{\hspace{1pt}}c@{\hspace{1pt}}}
$\stackrel{\hbox{\sffamily\scriptsize id}}{\downarrow}$ $\stackrel{\hbox{\sffamily\scriptsize step}}{\rightarrow}$&\tiny25&\tiny26&\tiny27&\tiny28&\tiny29&\tiny30&\tiny31&\tiny32&\tiny33&\tiny34&\tiny35&\tiny36&\tiny37&\tiny38&\tiny39&\tiny40&\tiny41&\tiny42&\tiny43&\tiny44&\tiny45&\tiny46&\tiny47\\ \hline
\sf 0&$-$&$-$&$-$&$R_{5}$&$-$&$-$&$-$&$-$&$-$&$R_{6}$&$-$&$-$&$-$&$-$&$-$&$-$&$R_{7}$&$-$&$-$&$-$&$-$&$-$&$-$\\
\sf 1&$-$&$-$&$-$&$-$&$R_{5}$&$-$&$-$&$-$&$-$&$-$&$R_{6}$&$-$&$-$&$-$&$-$&$-$&$-$&$R_{7}$&$-$&$-$&$-$&$-$&$-$\\
\sf 2&$R_{4}$&$-$&$-$&$-$&$-$&$R_{5}$&$-$&$-$&$-$&$-$&$-$&$R_{6}$&$-$&$-$&$-$&$-$&$-$&$-$&$R_{7}$&$-$&$-$&$-$&$-$\\
\sf 3&$S_{7}$&$R_{4}$&$-$&$-$&$-$&$-$&$R_{5}$&$-$&$-$&$-$&$-$&$-$&$R_{6}$&$-$&$-$&$-$&$-$&$-$&$-$&$R_{7}$&$-$&$-$&$-$\\
\sf 4&$S_{2}$&$S_{3}$&$S_{5}$&$S_{6}$&$S_{7}$&$-$&$-$&$R_{5}$&$-$&$-$&$-$&$-$&$-$&$R_{6}$&$-$&$-$&$-$&$-$&$-$&$-$&$R_{7}$&$-$&$-$\\
\sf 5&$-$&$-$&$R_{4}$&$S_{0}$&$S_{1}$&$S_{2}$&$S_{3}$&$S_{4}$&$S_{6}$&$S_{7}$&$-$&$-$&$-$&$-$&$R_{6}$&$-$&$-$&$-$&$-$&$-$&$-$&$R_{7}$&$-$\\
\sf 6&$-$&$-$&$-$&$R_{4}$&$-$&$-$&$-$&$-$&$R_{5}$&$S_{0}$&$S_{1}$&$S_{2}$&$S_{3}$&$S_{4}$&$S_{5}$&$S_{7}$&$-$&$-$&$-$&$-$&$-$&$-$&$R_{7}$\\
\sf 7&$R_{3}$&$-$&$-$&$-$&$R_{4}$&$-$&$-$&$-$&$-$&$R_{5}$&$-$&$-$&$-$&$-$&$-$&$R_{6}$&$S_{0}$&$S_{1}$&$S_{2}$&$S_{3}$&$S_{4}$&$S_{5}$&$S_{6}$\\
\hline
$\stackrel{\hbox{\sffamily\scriptsize $\vec\nu$}}{\rightarrow}$&4&2&2&4&4&2&2&2&2&4&2&2&2&2&2&2&2&2&2&2&2&2&2
\end{tabular}
\caption{Run-table 7 for $\mathcal P$ equal to the identity permutation.
Average utilization is 2.38 slots out of 8, or an efficiency of 29.79\%.}
\label{runtable7}
\end{table}

Finally, the removal of the last occurrence of $S_{N+1}$ from the
series of sending actions which constitute the broadcast of processor $N$
allows the removal of the whole right sub-table starting at that
point. The obtained table contains all and only
the actions of run $N$; the coherence
of these action is not affected; and all broadcast sessions are managed 
according to the rule of the identity permutation. In other words, this is
run-table $N$.

Now let us consider $\sigma(N+1)$: according to the above argument,
this is equal to:
\begin{enumerate}
\item the number of slots available in a
$N$-run i.e., $\sigma(N)$,
\item plus $N+1$ slots from each of the columns 
that witness a delay i.e.,  
\[ \floor{(N+1)/2}\cdot (N+1),\]
\item plus the slots in the right sub-matrix, not counting
the last row i.e., 
\[ (N+1)(N+2),\]
\item plus an additional row.
\end{enumerate}

In other words, $\sigma(N+1)$ can be expressed as the sum of the above
first three item multiplied by a factor equal to $\frac{N+2}{N+1}$.
This can be written as an equation as
\begin{equation}\label{left}
\sigma(k+1)= \bigl( \sigma(k) + \floor{\frac{k+1}2}(k+1) + (k+1)(k+2) \bigr)
\frac{k+2}{k+1}.
\end{equation}

By Definition~\ref{sigma}, this brings to the following recursive relation:
\begin{equation}\label{recurel}
\lambda(k+1) = \lambda(k) + \floor{\frac{k+1}2} + k + 2.
\end{equation}

Furthermore, the following is true by the induction hypothesis:
\begin{equation}\label{hyp}
\lambda_N = \lambda(N) = \frac34N^2 + \frac54N + \frac12 \floor{N/2},
\end{equation}

Our goal is to show that Eq.~(\ref{recurel}) and Eq.~(\ref{hyp})
together imply that $\lambda(N+1)=\lambda_{N+1}$, this latter being

\begin{eqnarray}\label{th1}
\lambda_{N+1} & = & \frac34(N+1)^2 + \frac54(N+1) + \frac12 \floor{\frac{N+1}{2}} \\
 \label{th2}  & = & \frac34N^2 + \frac{11}4N + 2 +  \frac12 \floor{\frac{N+1}{2}}.
\end{eqnarray}

Now, let us suppose $N$ is even---this implies that
$\floor{(N+1)/2} = \floor{N/2} = N/2$.
Exploiting this in Eq.~(\ref{recurel}) for $k=N$ and in Eq.~(\ref{hyp}), and
substituting the latter in the former Equation, brings us
to the following result:
\begin{eqnarray}
\lambda(N+1) &=& \lambda(N) + \floor{\frac{N+1}{2}} + N + 2 \nonumber\\
             &=& \lambda_N + \frac32N + 2 \nonumber\\
             &=& \frac34N^2 + \frac54N + \frac{N}4 + \frac32N + 2 \nonumber\\
             &=& \frac34N^2 + 3N + 2 \nonumber\\
             &=& \frac34N^2 + \frac{11}4N + 2 + \frac12 \floor{\frac{N+1}{2}},\nonumber
\end{eqnarray}

which is equal to $\lambda_{N+1}$ because of Eq.~(\ref{th2}). 
On the other hand, if $N>0$ is odd, then $\floor{(N+1)/2}=(N+1)/2$, while
$\floor{N/2}=(N-1)/2$. With the same approach as above we get:
\begin{eqnarray}
\lambda(N+1) &=& \lambda(N) + \floor{\frac{N+1}{2}} + N + 2 \nonumber\\
             &=& \frac34N^2 + \frac54N + \frac12 \floor{N/2} + (N+1)/2 + N + 2 \nonumber\\
             &=& \frac34N^2 + \frac54N + (N-1)/4 + (N+1)/2 + N + 2 \nonumber\\
	     &=& \frac34N^2 + \frac32N + \frac34 + \frac54N + \frac54+ (N+1)/4\nonumber\\
             &=& \frac34(N+1)^2 + \frac54(N+1) + \frac12\floor{\frac{N+1}{2}},\nonumber
\end{eqnarray}
which is again equal to $\lambda_{N+1}$ because of Eq.~(\ref{th1}). 
\qed
\end{pf*}

\begin{lem}\label{4slot.clusters}
The number of columns with 4 used slots inside, for a run with $\mathcal P$ equal to the identity 
permutation and $N+1$ processors, is
\begin{equation}
\sum_{i=0}^{N-1} \floor{\frac{i}{2}}.
\end{equation}
\end{lem}
\begin{pf}
Figure~\ref{f-prop1} shows also how, for any processor $1\leq i\leq\floor{(N+1)/2}$, there exists
only one cluster of $i-1$ columns such that each column contains exactly 4 used slots.
Moreover Fig.~\ref{f-prop2} shows that, for any processor $\floor{(N+1)/2}+1\leq i\leq N$, there exists
only one cluster of $N-i$ columns with that same property.

\begin{figure}[h]
\centerline{\psfig{figure=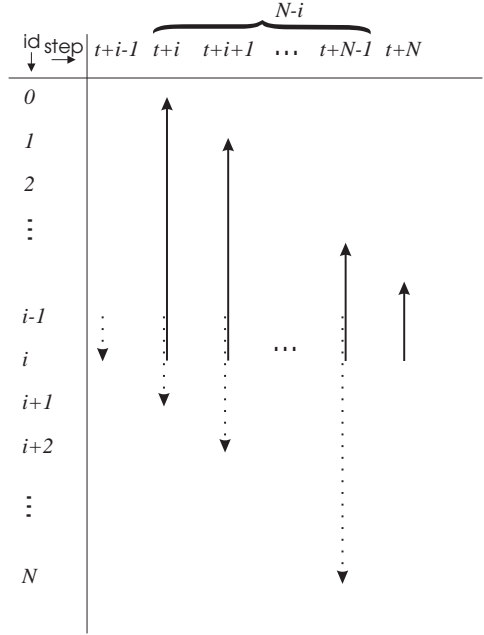,height=8.4cm}}
\caption{For any processor $i>\floor{(N+1)/2}$, there exists only one
cluster of $N-i$ columns with 4 used slots inside.}
\label{f-prop2}
\end{figure}

Let us call $u_4$ this number and count such columns:
\begin{equation}
u_4 = \sum_{i=1}^{\floor{(N+1)/2}} (i-1) + \sum_{i=\floor{(N+1)/2}+1}^N (N-i).
\end{equation}
Via two well-known algebraic transformations on sums (see
e.g., in~\cite{GrKn86,PuBr85})
we get to
\begin{equation}
u_4 = \sum_{j=0}^{\floor{(N+1)/2}-1} j + \sum_{j=0}^{N-\floor{(N+1)/2}-1} (N-\floor{\frac{N+1}2}-1-j).
\end{equation}

Now, if $N$ is even, then
\begin{eqnarray}
u_4 &=& \sum_{j=0}^{N/2-1} j + \sum_{j=0}^{N/2-1} (\frac{N}2-1-j) \nonumber\\
    &=& \sum_{j=0}^{N/2-1} (\frac{N}2-1) \nonumber\\
    &=& (\frac{N}2)(\frac{N}2-1) \label{even.quadratic} \\ 
    &=& 2\sum_{i=0}^{N/2-1} i \nonumber\\
    &=& \sum_{i=0}^{N-1} \floor{\frac{i}2},\nonumber
\end{eqnarray}

\noindent
while, if $N$ is odd,
\begin{eqnarray}
u_4 &=& \sum_{j=0}^{(N-1)/2} j + \sum_{j=0}^{(N-3)/2} (\frac{N-3}2-j) \nonumber\\
    &=& \frac{N-1}2 + \sum_{j=0}^{(N-3)/2} \frac{N-3}2 \nonumber\\
    &=& \frac{N-1}2 + \frac{N-1}2 \frac{N-3}2 \label{odd.quadratic} \\ 
    &=& \frac{N-1}2 + 2\sum_{i=0}^{(N-3)/2} i \nonumber\\
    &=& \sum_{i=0}^{N-1} \floor{\frac{i}2}.\nonumber
\end{eqnarray}
\qed
\end{pf}

\begin{figure}
\centerline{\psfig{figure=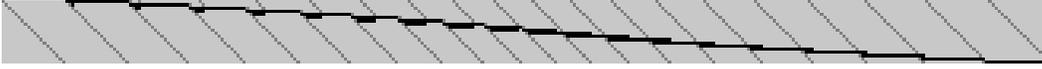,height=\textwidth,angle=-90}}
\caption{A graphical representation for run-table 20 when $\mathcal P$ is
the identity permutation. Light gray pixels represent wasted slots, 
gray pixels represent $R$ actions, black slots are sending actions.
Note the black ``blocks'' which represent the clusters mentioned 
in Fig.~\ref{f-prop1} and Fig.~\ref{f-prop2}.}
\label{z20}
\end{figure}

Figure~\ref{z20} shows the typical shape of run-tables 
in the case of $\mathcal P$ being the identity permutation, also
locating the 4-used slot clusters. 


The following Propositions locate the asymptotic values of $\mu$ and $\varepsilon$:
\begin{prop}\label{id-eps}
$\lim_{k\rightarrow\infty}\varepsilon_k=0.$
\end{prop}
\begin{pf}
Let us call $U(k)$ the number of used slots in a run of $k$ processors.
As a consequence of Lemma~\ref{4slot.clusters}, the number of used slots in a run is
\begin{eqnarray}
U(k) &=& 2\sum_{i=0}^{k-1} \floor{\frac{i}{2}} + 2 \lambda_k \nonumber\\
     &=& 2\sum_{i=0}^{k-1} \floor{\frac{i}{2}} + 2 (\frac34k^2 + \frac54k + \frac12\floor{k/2}).
\end{eqnarray}
From Definition~\ref{sigma} we derive that
\begin{equation}\label{epsilon}
\varepsilon_k = \frac{U(k)}{(k+1)\lambda_k}.
\end{equation}
Eq.~(\ref{even.quadratic}) and Eq.~(\ref{odd.quadratic}) show that 
$\deg [U(k)]=2$,
while from Prop.~\ref{prop1} we know that $\deg [(k+1)\cdot\lambda_k] = 3$.
As a consequence, $\varepsilon_k$ tends to zero as $k$ tends to infinity.
\qed
\end{pf}

\begin{prop}\label{id-mu}
$\lim_{k\rightarrow\infty}\mu_k=\frac{8}3.$
\end{prop}
\begin{pf}
Being 
\begin{equation}\label{mu}
\mu_k = \frac{U(k)}{\lambda_k},
\end{equation}
it is possible to derive that
\begin{eqnarray}
\mu_k &=& \frac{2\frac{k^2}4 + 2 \frac34k^2 + \hbox{ \emph{ \ldots some 1st degree elements}}}%
               {\frac34k^2 + \hbox{ \emph{ \ldots some 1st degree elements}}}\nonumber \\
      &=& \frac{2k^2 + \hbox{ \emph{ \ldots some 1st degree elements}}}%
               {\frac34k^2 + \hbox{ \emph{ \ldots some 1st degree elements}}},
\end{eqnarray}
which tends to $\frac{8}3$, or $2.\overline{6}$, when $k$ goes to infinity.
\qed
\end{pf}

\section{Second Case: Pseudo-random Permutations}\label{randcase}
This Section covers the case such that
$\mathcal P$ is a pseudo-random\footnote{%
	The standard C function ``{\tt random}''~\cite{Rand} has been 
	used---a non-linear additive feedback random number
	generator returning pseudo-random numbers in the range $[0,2^{31}-1]$
	with a period approximately equal to $16(2^{31}-1)$. A truly
	random integer has been used as a seed.}
permutation of the integers $0,\dots,i-1,i+1,\dots,N$.

\begin{figure}
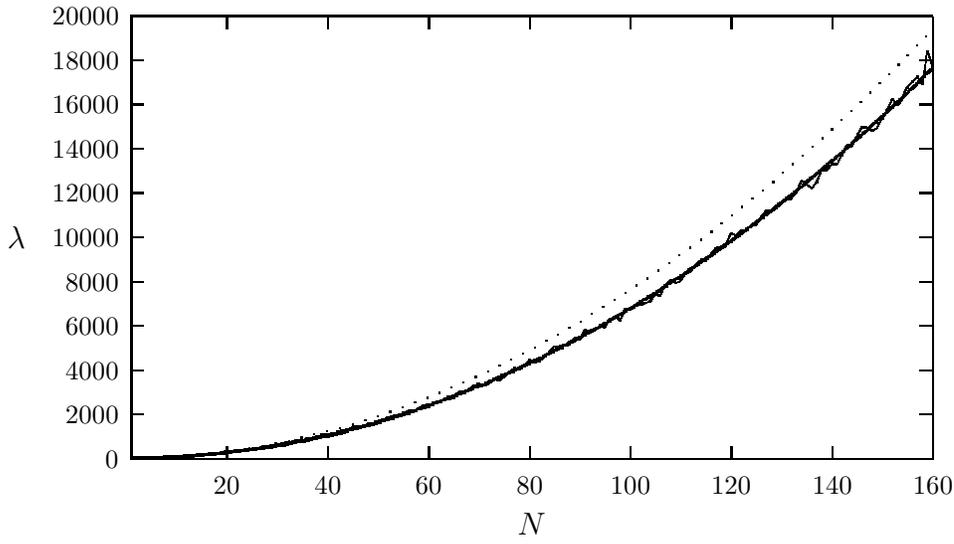

\input cmpsteps-zerorand
\caption{Comparison between lengths in the case of the identity
permutation (dotted parabola) and that of the random permutation
(piecewise line), $1\leq N\leq160$. The lowest curve
($\lambda=0.71 N^2-3.88 N + 88.91$) is the parabola
best fitting with the piecewise line---which suggests
a quadratic execution time as in the case of the identity permutation.}
\label{cmpsteps-zerorand}
\end{figure}

\begin{figure}
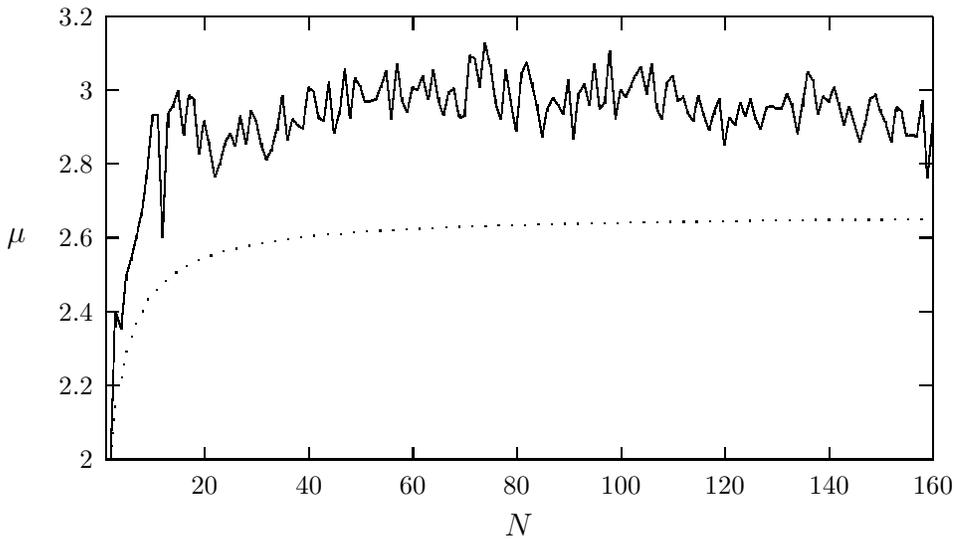

\input cmpaverage-zerorand
\caption{Comparison between values of $\mu$ in the case of 
a pseudo-random permutation (piecewise line) and that of the
identity permutation (dotted curve), $1\leq N\leq160$. Note how the former is
strictly over the latter. Note also how $\mu$ seems to tend to
a value right above $2.6$ for the identity permutation,
as claimed by Prop.~\ref{id-mu}.}
\label{cmpaverage-zerorand}
\end{figure}

Figure~\ref{cmpsteps-zerorand} shows the values of $\lambda$
using the identity and random permutations
and graphs the parabola who best fits with these latter values.
We conclude that, experimentally, the choice of case one is 
even ``worse'' than choosing permutations at random.
The same conclusion follows from
Fig.~\ref{cmpaverage-zerorand} and Fig.~\ref{cmpefficiency-zerorand}
which respectively confront the averages and efficiencies
in the above two cases.

\begin{figure}
\centerline{\psfig{figure=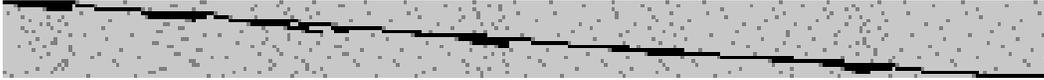,height=\textwidth,angle=-90}}
\caption{A graphical representation for run-table 20 when $\mathcal P$ is
a pseudo-random permutation.}\label{r20}
\end{figure}

Table~\ref{rantable} shows run-table 5, and
Fig.~\ref{r20} shows the shape of run-table 20 in this case.

\begin{table}
\begin{tabular}{l|c@{\hspace{1pt}}c@{\hspace{1pt}}c@{\hspace{1pt}}c@{\hspace{1pt}}c@{\hspace{1pt}}c@{\hspace{1pt}}c@{\hspace{1pt}}c@{\hspace{1pt}}c@{\hspace{1pt}}c@{\hspace{1pt}}c@{\hspace{1pt}}c@{\hspace{1pt}}c@{\hspace{1pt}}c@{\hspace{1pt}}c@{\hspace{1pt}}c@{\hspace{1pt}}c@{\hspace{1pt}}c@{\hspace{1pt}}c@{\hspace{1pt}}c@{\hspace{1pt}}c@{\hspace{1pt}}c@{\hspace{1pt}}c@{\hspace{1pt}}c@{\hspace{1pt}}}
$\stackrel{\hbox{\sffamily\scriptsize id}}{\downarrow}$ $\stackrel{\hbox{\sffamily\scriptsize step}}{\rightarrow}$&\tiny1&\tiny2&\tiny3&\tiny4&\tiny5&\tiny6&\tiny7&\tiny8&\tiny9&\tiny10&\tiny11&\tiny12&\tiny13&\tiny14&\tiny15&\tiny16&\tiny17&\tiny18&\tiny
19&\tiny20&\tiny21&\tiny22&\tiny23&\tiny24\\ \hline
\sf 0&$S_{5}$&$S_{1}$&$S_{3}$&$S_{2}$&$S_{4}$&$R_{1}$&$-$&$-$&$-$&$-$&$R_{2}$&$-$&$-$&$-$&$R_{3}$&$-$&$-$&$-$&$-$&$R_{4}$&$R_{5}$&$-$&$-$&$-$\\
\sf 1&$-$&$R_{0}$&$S_{5}$&$\curvearrowright$&$\curvearrowright$&$S_{0}$&$S_{3}$&$S_{2}$&$S_{4}$&$R_{2}$&$-$&$-$&$-$&$R_{3}$&$-$&$-$&$-$&$-$&$R_{4}$&$R_{5}$&$-$&$-$&$-$&$-$\\
\sf 2&$-$&$-$&$-$&$R_{0}$&$-$&$-$&$-$&$R_{1}$&$S_{5}$&$S_{1}$&$S_{0}$&$S_{3}$&$S_{4}$&$-$&$-$&$R_{3}$&$-$&$-$&$-$&$-$&$-$&$R_{4}$&$R_{5}$&$-$\\
\sf 3&$-$&$-$&$R_{0}$&$-$&$-$&$-$&$R_{1}$&$-$&$-$&$-$&$-$&$R_{2}$&$S_{5}$&$S_{1}$&$S_{0}$&$S_{2}$&$S_{4}$&$-$&$-$&$-$&$R_{4}$&$R_{5}$&$-$&$-$\\
\sf 4&$-$&$-$&$-$&$-$&$R_{0}$&$-$&$-$&$-$&$R_{1}$&$-$&$-$&$-$&$R_{2}$&$-$&$-$&$-$&$R_{3}$&$S_{5}$&$S_{1}$&$S_{0}$&$S_{3}$&$S_{2}$&$-$&$R_{5}$\\
\sf 5&$R_{0}$&$-$&$R_{1}$&$-$&$-$&$-$&$-$&$-$&$R_{2}$&$-$&$-$&$-$&$R_{3}$&$-$&$-$&$-$&$-$&$R_{4}$&$\curvearrowright$&$S_{1}$&$S_{0}$&$S_{3}$&$S_{2}$&$S_{4}$\\
\hline
$\stackrel{\hbox{\sffamily\scriptsize $\vec\nu$}}{\rightarrow}$&2&2&4&2&2&2&2&2&4&2&2&2&4&2&2&2&2&2&2&4&4&4&2&2
\end{tabular}
\caption{Run-table 5 when $\mathcal P$ is chosen pseudo-randomly.
$\mu$ is $2.5$ slots out of 6, which implies an efficiency of 41.67\%.}
\label{rantable}
\end{table}

\begin{figure}
\input cmpefficiency-zerorand
\caption{Comparison between values of $\varepsilon$ in the case of 
the random permutation (piecewise line) and that of the
identity permutation (dotted curve), $1\leq N\leq160$. 
Also in this graph the former is strictly over the latter, though they get
closer to each other and to zero as $N$ increases, as 
proven for the identity permutation in Prop.~\ref{id-eps}.}
\label{cmpefficiency-zerorand}
\end{figure}

\section{Third Case: the Algorithm of Pipelined Broadcast}\label{tao-pb}
Let $\mathcal P$ be the following permutation:
\begin{equation}\label{pipe}
\binom{0,\dots,i-1,i+1,\dots,N}%
      {i+1,\dots,N,0,\dots,i-1}.
\end{equation}

Note how permutation~(\ref{pipe}) is equivalent to $i$ 
cyclic logical left shifts of the identity permutation. Note also how,
in cycle notation~\cite{Knu73a}, (\ref{pipe}) is represented as one cycle;
for instance,

\[
\binom{0,1,2,4,5}%
      {4,5,0,1,2},
\]

i.e., (\ref{pipe}) for $N=5$ and $i=3$, is equivalent to cycle $(0,4,1,5,2)$.

A value of $\mathcal P$ equal to permutation~(\ref{pipe}) means that, once processor $i$ 
has gained the right to broadcast,
it will first send its message to processor $i+1$
(possibly having to wait for it to become available to receive that message),
then it will do the same with processor $i+2$, and so
forth up to $N$, then wrapping around and going from processor 0 to processor $i-1$.
This is represented in Table~\ref{run9} for $N=9$.

Pictures quite similar to Table~\ref{run9} can be found in many classical works
on pipelined microprocessors (see e.g.~\cite[p.132--133]{PaHe96}.)
Indeed, a pipeline is a series of data-paths shifted in time so to overlap
their execution, the same way Eq.~(\ref{pipe}) tends to overlap
as much as possible its broadcast sessions. Clearly pipe stages are represented
here as full processors, and the concept of machine cycle, or pipe stage time
of pipelined processor, simply collapses to the concept of time step as
introduced in Def.~\ref{timestep}.

A number of considerations like those above brought us to the name
we use for this special case of our algorithm, as algorithm of
``pipelined gossiping.'' We will remark them in the following
using the italics typeface.


\begin{table}
\begin{tabular}{l|c@{\hspace{1pt}}c@{\hspace{1pt}}c@{\hspace{1pt}}c@{\hspace{1pt}}c@{\hspace{1pt}}c@{\hspace{1pt}}c@{\hspace{1pt}}c@{\hspace{1pt}}c@{\hspace{1pt}}c@{\hspace{1pt}}c@{\hspace{1pt}}c@{\hspace{1pt}}c@{\hspace{1pt}}c@{\hspace{1pt}}c@{\hspace{1pt}}c@{\hspace{1pt}}c@{\hspace{1pt}}c@{\hspace{1pt}}c@{\hspace{1pt}}c@{\hspace{1pt}}c@{\hspace{1pt}}c@{\hspace{1pt}}c@{\hspace{1pt}}c@{\hspace{1pt}}c@{\hspace{1pt}}c@{\hspace{1pt}}c@{\hspace{1pt}}}
$\stackrel{\hbox{\sffamily\scriptsize id}}{\downarrow}$ $\stackrel{\hbox{\sffamily\scriptsize step}}{\rightarrow}$&\tiny1&\tiny2&\tiny3&\tiny4&\tiny5&\tiny6&\tiny7&\tiny8&\tiny9&\tiny10&\tiny11&\tiny12&\tiny13&\tiny14&\tiny15&\tiny16&\tiny17&\tiny18&\tiny
19&\tiny20&\tiny21&\tiny22&\tiny23&\tiny24&\tiny25&\tiny26&\tiny27\\ \hline
\sf 0&$S_{1}$&$S_{2}$&$S_{3}$&$S_{4}$&$S_{5}$&$S_{6}$&$S_{7}$&$S_{8}$&$S_{9}$&$-$&$R_{1}$&$R_{2}$&$R_{3}$&$R_{4}$&$R_{5}$&$R_{6}$&$R_{7}$&$R_{8}$&$R_{9}$&$-$&$-$&$-$&$-$&$-$&$-$&$-$&$-$\\
\sf 1&$R_{0}$&$\curvearrowright$&$S_{2}$&$S_{3}$&$S_{4}$&$S_{5}$&$S_{6}$&$S_{7}$&$S_{8}$&$S_{9}$&$S_{0}$&$-$&$R_{2}$&$R_{3}$&$R_{4}$&$R_{5}$&$R_{6}$&$R_{7}$&$R_{8}$&$R_{9}$&$-$&$-$&$-$&$-$&$-$&$-$&$-$\\
\sf 2&$-$&$R_{0}$&$R_{1}$&$\curvearrowright$&$S_{3}$&$S_{4}$&$S_{5}$&$S_{6}$&$S_{7}$&$S_{8}$&$S_{9}$&$S_{0}$&$S_{1}$&$-$&$R_{3}$&$R_{4}$&$R_{5}$&$R_{6}$&$R_{7}$&$R_{8}$&$R_{9}$&$-$&$-$&$-$&$-$&$-$&$-$\\
\sf 3&$-$&$-$&$R_{0}$&$R_{1}$&$R_{2}$&$\curvearrowright$&$S_{4}$&$S_{5}$&$S_{6}$&$S_{7}$&$S_{8}$&$S_{9}$&$S_{0}$&$S_{1}$&$S_{2}$&$-$&$R_{4}$&$R_{5}$&$R_{6}$&$R_{7}$&$R_{8}$&$R_{9}$&$-$&$-$&$-$&$-$&$-$\\
\sf 4&$-$&$-$&$-$&$R_{0}$&$R_{1}$&$R_{2}$&$R_{3}$&$\curvearrowright$&$S_{5}$&$S_{6}$&$S_{7}$&$S_{8}$&$S_{9}$&$S_{0}$&$S_{1}$&$S_{2}$&$S_{3}$&$-$&$R_{5}$&$R_{6}$&$R_{7}$&$R_{8}$&$R_{9}$&$-$&$-$&$-$&$-$\\
\sf 5&$-$&$-$&$-$&$-$&$R_{0}$&$R_{1}$&$R_{2}$&$R_{3}$&$R_{4}$&$\curvearrowright$&$S_{6}$&$S_{7}$&$S_{8}$&$S_{9}$&$S_{0}$&$S_{1}$&$S_{2}$&$S_{3}$&$S_{4}$&$-$&$R_{6}$&$R_{7}$&$R_{8}$&$R_{9}$&$-$&$-$&$-$\\
\sf 6&$-$&$-$&$-$&$-$&$-$&$R_{0}$&$R_{1}$&$R_{2}$&$R_{3}$&$R_{4}$&$R_{5}$&$\curvearrowright$&$S_{7}$&$S_{8}$&$S_{9}$&$S_{0}$&$S_{1}$&$S_{2}$&$S_{3}$&$S_{4}$&$S_{5}$&$-$&$R_{7}$&$R_{8}$&$R_{9}$&$-$&$-$\\
\sf 7&$-$&$-$&$-$&$-$&$-$&$-$&$R_{0}$&$R_{1}$&$R_{2}$&$R_{3}$&$R_{4}$&$R_{5}$&$R_{6}$&$\curvearrowright$&$S_{8}$&$S_{9}$&$S_{0}$&$S_{1}$&$S_{2}$&$S_{3}$&$S_{4}$&$S_{5}$&$S_{6}$&$-$&$R_{8}$&$R_{9}$&$-$\\
\sf 8&$-$&$-$&$-$&$-$&$-$&$-$&$-$&$R_{0}$&$R_{1}$&$R_{2}$&$R_{3}$&$R_{4}$&$R_{5}$&$R_{6}$&$R_{7}$&$\curvearrowright$&$S_{9}$&$S_{0}$&$S_{1}$&$S_{2}$&$S_{3}$&$S_{4}$&$S_{5}$&$S_{6}$&$S_{7}$&$-$&$R_{9}$\\
\sf 9&$-$&$-$&$-$&$-$&$-$&$-$&$-$&$-$&$R_{0}$&$R_{1}$&$R_{2}$&$R_{3}$&$R_{4}$&$R_{5}$&$R_{6}$&$R_{7}$&$R_{8}$&$\curvearrowright$&$S_{0}$&$S_{1}$&$S_{2}$&$S_{3}$&$S_{4}$&$S_{5}$&$S_{6}$&$S_{7}$&$S_{8}$\\
\hline
$\stackrel{\hbox{\sffamily\scriptsize $\vec\nu$}}{\rightarrow}$&2&2&4&4&6&6&8&8&10&8&10&8&10&8&10&8&10&8&10&8&8&6&6&4&4&2&2
\end{tabular}
\caption{Run-table of a run for $N=9$ using permutation of Eq.~(\ref{pipe}).
In this case $\mu$, or the average utilization is 6.67 slots out of 10, 
with an efficiency $\varepsilon=66.67\%$ and a length $\lambda=27$.
Note that $\vec\nu$ is in this case a palindrome i.e., 
as well known~\cite{Vil71},
a string like ``21012'' which can be read indifferently from left to right 
or vice-versa.}\label{run9}
\end{table}

Clearly using this permutation leads to better performance. In particular,
after a start-up phase (\emph{after filling the pipeline\/}),
sustained performance
is close to the maximum---a number of unused slots 
(\emph{pipeline bubbles\/}) still exist, even
in the sustained region, but here $\mu$ reaches value $N+1$ half 
of the times (if $N$ is odd).
In the region of decay, starting from time step 19, every new time step a processor
fully completes its task. Similar remarks apply to Table~\ref{run8};
this is
the typical shape of a run-table for $N$ even. This time the state within the
sustained region is more steady, though the maximum number of used slots
never reaches the number of slots in the system.

\begin{table}
\begin{tabular}{l|c@{\hspace{1pt}}c@{\hspace{1pt}}c@{\hspace{1pt}}c@{\hspace{1pt}}c@{\hspace{1pt}}c@{\hspace{1pt}}c@{\hspace{1pt}}c@{\hspace{1pt}}c@{\hspace{1pt}}c@{\hspace{1pt}}c@{\hspace{1pt}}c@{\hspace{1pt}}c@{\hspace{1pt}}c@{\hspace{1pt}}c@{\hspace{1pt}}c@{\hspace{1pt}}c@{\hspace{1pt}}c@{\hspace{1pt}}c@{\hspace{1pt}}c@{\hspace{1pt}}c@{\hspace{1pt}}c@{\hspace{1pt}}c@{\hspace{1pt}}c@{\hspace{1pt}}}
$\stackrel{\hbox{\sffamily\scriptsize id}}{\downarrow}$ $\stackrel{\hbox{\sffamily\scriptsize step}}{\rightarrow}$&\tiny1&\tiny2&\tiny3&\tiny4&\tiny5&\tiny6&\tiny7&\tiny8&\tiny9&\tiny10&\tiny11&\tiny12&\tiny13&\tiny14&\tiny15&\tiny16&\tiny17&\tiny18&\tiny
19&\tiny20&\tiny21&\tiny22&\tiny23&\tiny24\\ \hline
\sf 0&$S_{1}$&$S_{2}$&$S_{3}$&$S_{4}$&$S_{5}$&$S_{6}$&$S_{7}$&$S_{8}$&$-$&$R_{1}$&$R_{2}$&$R_{3}$&$R_{4}$&$R_{5}$&$R_{6}$&$R_{7}$&$R_{8}$&$-$&$-$&$-$&$-$&$-$&$-$&$-$\\
\sf 1&$R_{0}$&$\curvearrowright$&$S_{2}$&$S_{3}$&$S_{4}$&$S_{5}$&$S_{6}$&$S_{7}$&$S_{8}$&$S_{0}$&$-$&$R_{2}$&$R_{3}$&$R_{4}$&$R_{5}$&$R_{6}$&$R_{7}$&$R_{8}$&$-$&$-$&$-$&$-$&$-$&$-$\\
\sf 2&$-$&$R_{0}$&$R_{1}$&$\curvearrowright$&$S_{3}$&$S_{4}$&$S_{5}$&$S_{6}$&$S_{7}$&$S_{8}$&$S_{0}$&$S_{1}$&$-$&$R_{3}$&$R_{4}$&$R_{5}$&$R_{6}$&$R_{7}$&$R_{8}$&$-$&$-$&$-$&$-$&$-$\\
\sf 3&$-$&$-$&$R_{0}$&$R_{1}$&$R_{2}$&$\curvearrowright$&$S_{4}$&$S_{5}$&$S_{6}$&$S_{7}$&$S_{8}$&$S_{0}$&$S_{1}$&$S_{2}$&$-$&$R_{4}$&$R_{5}$&$R_{6}$&$R_{7}$&$R_{8}$&$-$&$-$&$-$&$-$\\
\sf 4&$-$&$-$&$-$&$R_{0}$&$R_{1}$&$R_{2}$&$R_{3}$&$\curvearrowright$&$S_{5}$&$S_{6}$&$S_{7}$&$S_{8}$&$S_{0}$&$S_{1}$&$S_{2}$&$S_{3}$&$-$&$R_{5}$&$R_{6}$&$R_{7}$&$R_{8}$&$-$&$-$&$-$\\
\sf 5&$-$&$-$&$-$&$-$&$R_{0}$&$R_{1}$&$R_{2}$&$R_{3}$&$R_{4}$&$\curvearrowright$&$S_{6}$&$S_{7}$&$S_{8}$&$S_{0}$&$S_{1}$&$S_{2}$&$S_{3}$&$S_{4}$&$-$&$R_{6}$&$R_{7}$&$R_{8}$&$-$&$-$\\
\sf 6&$-$&$-$&$-$&$-$&$-$&$R_{0}$&$R_{1}$&$R_{2}$&$R_{3}$&$R_{4}$&$R_{5}$&$\curvearrowright$&$S_{7}$&$S_{8}$&$S_{0}$&$S_{1}$&$S_{2}$&$S_{3}$&$S_{4}$&$S_{5}$&$-$&$R_{7}$&$R_{8}$&$-$\\
\sf 7&$-$&$-$&$-$&$-$&$-$&$-$&$R_{0}$&$R_{1}$&$R_{2}$&$R_{3}$&$R_{4}$&$R_{5}$&$R_{6}$&$\curvearrowright$&$S_{8}$&$S_{0}$&$S_{1}$&$S_{2}$&$S_{3}$&$S_{4}$&$S_{5}$&$S_{6}$&$-$&$R_{8}$\\
\sf 8&$-$&$-$&$-$&$-$&$-$&$-$&$-$&$R_{0}$&$R_{1}$&$R_{2}$&$R_{3}$&$R_{4}$&$R_{5}$&$R_{6}$&$R_{7}$&$\curvearrowright$&$S_{0}$&$S_{1}$&$S_{2}$&$S_{3}$&$S_{4}$&$S_{5}$&$S_{6}$&$S_{7}$\\
\hline
$\stackrel{\hbox{\sffamily\scriptsize $\vec\nu$}}{\rightarrow}$&2&2&4&4&6&6&8&8&8&8&8&8&8&8&8&8&8&8&6&6&4&4&2&2
\end{tabular}
\caption{Run-table of a run for $N=8$ using the permutation of Eq.~(\ref{pipe}).
$\mu$ is equal to 6 slots out of 9, with an efficiency $\varepsilon=66.67\%$ and a length $\lambda=24$.
Note how $\vec\nu$ is a palindrome string.%
}
\label{run8}
\end{table}

\begin{figure}
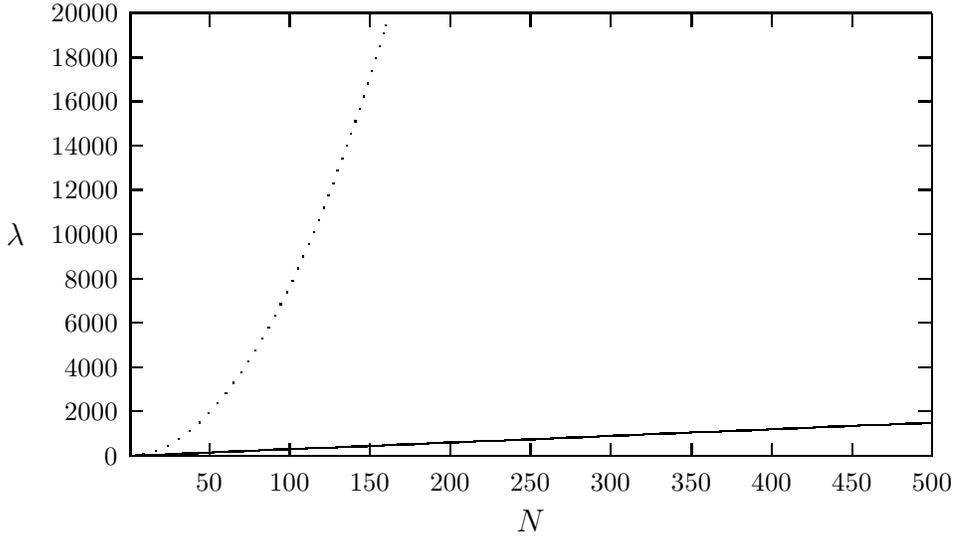

\input cmp12-steps
\caption{Comparison between run lengths resulting from 
the identity permutation (dotted parabola)
and those from permutation (\ref{pipe}). The former are shown for 
$1\leq N\leq 160$, the latter for $1\leq N\leq 500$.}
\label{cmp12-steps}
\end{figure}

\begin{figure}
\setlength{\unitlength}{0.240900pt}
\ifx\plotpoint\undefined\newsavebox{\plotpoint}\fi
\sbox{\plotpoint}{\rule[-0.200pt]{0.400pt}{0.400pt}}%
\begin{picture}(1500,900)(0,0)
\font\gnuplot=cmr10 at 10pt
\gnuplot
\sbox{\plotpoint}{\rule[-0.200pt]{0.400pt}{0.400pt}}%
\put(141.0,163.0){\rule[-0.200pt]{4.818pt}{0.400pt}}
\put(121,163){\makebox(0,0)[r]{0}}
\put(1460.0,163.0){\rule[-0.200pt]{4.818pt}{0.400pt}}
\put(141.0,250.0){\rule[-0.200pt]{4.818pt}{0.400pt}}
\put(121,250){\makebox(0,0)[r]{1}}
\put(1460.0,250.0){\rule[-0.200pt]{4.818pt}{0.400pt}}
\put(141.0,337.0){\rule[-0.200pt]{4.818pt}{0.400pt}}
\put(121,337){\makebox(0,0)[r]{2}}
\put(1460.0,337.0){\rule[-0.200pt]{4.818pt}{0.400pt}}
\put(141.0,424.0){\rule[-0.200pt]{4.818pt}{0.400pt}}
\put(121,424){\makebox(0,0)[r]{3}}
\put(1460.0,424.0){\rule[-0.200pt]{4.818pt}{0.400pt}}
\put(141.0,511.0){\rule[-0.200pt]{4.818pt}{0.400pt}}
\put(121,511){\makebox(0,0)[r]{4}}
\put(1460.0,511.0){\rule[-0.200pt]{4.818pt}{0.400pt}}
\put(141.0,598.0){\rule[-0.200pt]{4.818pt}{0.400pt}}
\put(121,598){\makebox(0,0)[r]{5}}
\put(1460.0,598.0){\rule[-0.200pt]{4.818pt}{0.400pt}}
\put(141.0,685.0){\rule[-0.200pt]{4.818pt}{0.400pt}}
\put(121,685){\makebox(0,0)[r]{6}}
\put(1460.0,685.0){\rule[-0.200pt]{4.818pt}{0.400pt}}
\put(141.0,772.0){\rule[-0.200pt]{4.818pt}{0.400pt}}
\put(121,772){\makebox(0,0)[r]{7}}
\put(1460.0,772.0){\rule[-0.200pt]{4.818pt}{0.400pt}}
\put(141.0,859.0){\rule[-0.200pt]{4.818pt}{0.400pt}}
\put(121,859){\makebox(0,0)[r]{8}}
\put(1460.0,859.0){\rule[-0.200pt]{4.818pt}{0.400pt}}
\put(141.0,163.0){\rule[-0.200pt]{0.400pt}{4.818pt}}
\put(141,122){\makebox(0,0){1}}
\put(141.0,839.0){\rule[-0.200pt]{0.400pt}{4.818pt}}
\put(290.0,163.0){\rule[-0.200pt]{0.400pt}{4.818pt}}
\put(290,122){\makebox(0,0){2}}
\put(290.0,839.0){\rule[-0.200pt]{0.400pt}{4.818pt}}
\put(439.0,163.0){\rule[-0.200pt]{0.400pt}{4.818pt}}
\put(439,122){\makebox(0,0){3}}
\put(439.0,839.0){\rule[-0.200pt]{0.400pt}{4.818pt}}
\put(587.0,163.0){\rule[-0.200pt]{0.400pt}{4.818pt}}
\put(587,122){\makebox(0,0){4}}
\put(587.0,839.0){\rule[-0.200pt]{0.400pt}{4.818pt}}
\put(736.0,163.0){\rule[-0.200pt]{0.400pt}{4.818pt}}
\put(736,122){\makebox(0,0){5}}
\put(736.0,839.0){\rule[-0.200pt]{0.400pt}{4.818pt}}
\put(885.0,163.0){\rule[-0.200pt]{0.400pt}{4.818pt}}
\put(885,122){\makebox(0,0){6}}
\put(885.0,839.0){\rule[-0.200pt]{0.400pt}{4.818pt}}
\put(1034.0,163.0){\rule[-0.200pt]{0.400pt}{4.818pt}}
\put(1034,122){\makebox(0,0){7}}
\put(1034.0,839.0){\rule[-0.200pt]{0.400pt}{4.818pt}}
\put(1182.0,163.0){\rule[-0.200pt]{0.400pt}{4.818pt}}
\put(1182,122){\makebox(0,0){8}}
\put(1182.0,839.0){\rule[-0.200pt]{0.400pt}{4.818pt}}
\put(1331.0,163.0){\rule[-0.200pt]{0.400pt}{4.818pt}}
\put(1331,122){\makebox(0,0){9}}
\put(1331.0,839.0){\rule[-0.200pt]{0.400pt}{4.818pt}}
\put(1480.0,163.0){\rule[-0.200pt]{0.400pt}{4.818pt}}
\put(1480,122){\makebox(0,0){10}}
\put(1480.0,839.0){\rule[-0.200pt]{0.400pt}{4.818pt}}
\put(141.0,163.0){\rule[-0.200pt]{322.565pt}{0.400pt}}
\put(1480.0,163.0){\rule[-0.200pt]{0.400pt}{167.666pt}}
\put(141.0,859.0){\rule[-0.200pt]{322.565pt}{0.400pt}}
\put(41,511){\makebox(0,0){$\mu$}}
\put(810,61){\makebox(0,0){$N$}}
\put(141.0,163.0){\rule[-0.200pt]{0.400pt}{167.666pt}}
\put(141,337){\usebox{\plotpoint}}
\multiput(290.00,337.58)(1.289,0.499){113}{\rule{1.128pt}{0.120pt}}
\multiput(290.00,336.17)(146.660,58.000){2}{\rule{0.564pt}{0.400pt}}
\multiput(439.00,395.58)(1.280,0.499){113}{\rule{1.121pt}{0.120pt}}
\multiput(439.00,394.17)(145.674,58.000){2}{\rule{0.560pt}{0.400pt}}
\multiput(587.00,453.58)(1.289,0.499){113}{\rule{1.128pt}{0.120pt}}
\multiput(587.00,452.17)(146.660,58.000){2}{\rule{0.564pt}{0.400pt}}
\multiput(736.00,511.58)(1.289,0.499){113}{\rule{1.128pt}{0.120pt}}
\multiput(736.00,510.17)(146.660,58.000){2}{\rule{0.564pt}{0.400pt}}
\multiput(885.00,569.58)(1.289,0.499){113}{\rule{1.128pt}{0.120pt}}
\multiput(885.00,568.17)(146.660,58.000){2}{\rule{0.564pt}{0.400pt}}
\multiput(1034.00,627.58)(1.280,0.499){113}{\rule{1.121pt}{0.120pt}}
\multiput(1034.00,626.17)(145.674,58.000){2}{\rule{0.560pt}{0.400pt}}
\multiput(1182.00,685.58)(1.289,0.499){113}{\rule{1.128pt}{0.120pt}}
\multiput(1182.00,684.17)(146.660,58.000){2}{\rule{0.564pt}{0.400pt}}
\multiput(1331.00,743.58)(1.289,0.499){113}{\rule{1.128pt}{0.120pt}}
\multiput(1331.00,742.17)(146.660,58.000){2}{\rule{0.564pt}{0.400pt}}
\put(141.0,337.0){\rule[-0.200pt]{35.894pt}{0.400pt}}
\put(1480,801){\usebox{\plotpoint}}
\put(141,337){\usebox{\plotpoint}}
\multiput(141,337)(20.756,0.000){8}{\usebox{\plotpoint}}
\multiput(290,337)(20.637,2.216){7}{\usebox{\plotpoint}}
\multiput(439,353)(20.751,0.421){7}{\usebox{\plotpoint}}
\multiput(587,356)(20.726,1.113){7}{\usebox{\plotpoint}}
\multiput(736,364)(20.754,0.279){7}{\usebox{\plotpoint}}
\multiput(885,366)(20.748,0.557){8}{\usebox{\plotpoint}}
\multiput(1034,370)(20.754,0.280){7}{\usebox{\plotpoint}}
\multiput(1182,372)(20.751,0.418){7}{\usebox{\plotpoint}}
\multiput(1331,375)(20.755,0.139){7}{\usebox{\plotpoint}}
\put(1480,376){\usebox{\plotpoint}}
\end{picture}
\caption{Comparison between values of $\mu$ derived from
the identity permutation (dotted parabola)
and those from permutation (\ref{pipe}) for $1\leq N\leq10$.}
\label{cmp12-average}
\end{figure}
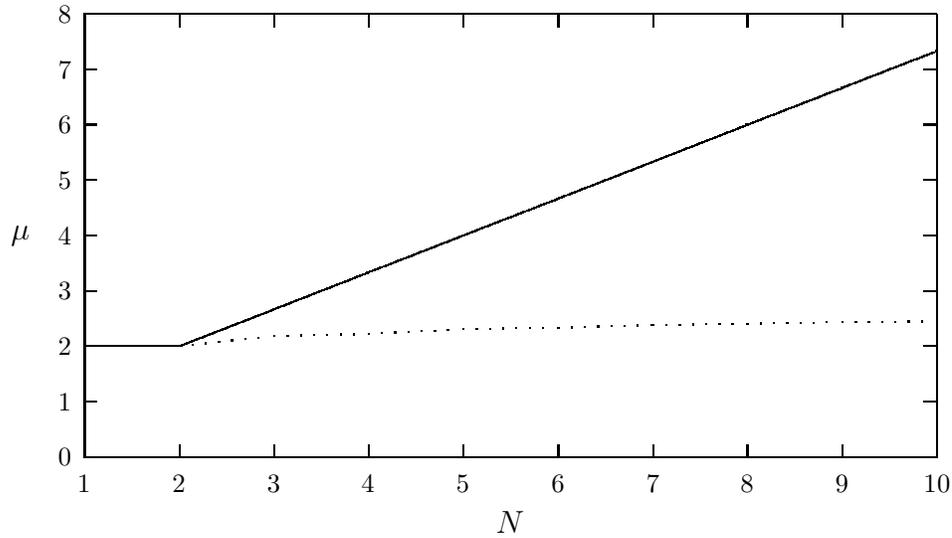

\begin{figure}
\setlength{\unitlength}{0.240900pt}
\ifx\plotpoint\undefined\newsavebox{\plotpoint}\fi
\sbox{\plotpoint}{\rule[-0.200pt]{0.400pt}{0.400pt}}%
\begin{picture}(1500,900)(0,0)
\font\gnuplot=cmr10 at 10pt
\gnuplot
\sbox{\plotpoint}{\rule[-0.200pt]{0.400pt}{0.400pt}}%
\put(181.0,163.0){\rule[-0.200pt]{4.818pt}{0.400pt}}
\put(161,163){\makebox(0,0)[r]{0}}
\put(1460.0,163.0){\rule[-0.200pt]{4.818pt}{0.400pt}}
\put(181.0,233.0){\rule[-0.200pt]{4.818pt}{0.400pt}}
\put(161,233){\makebox(0,0)[r]{10}}
\put(1460.0,233.0){\rule[-0.200pt]{4.818pt}{0.400pt}}
\put(181.0,302.0){\rule[-0.200pt]{4.818pt}{0.400pt}}
\put(161,302){\makebox(0,0)[r]{20}}
\put(1460.0,302.0){\rule[-0.200pt]{4.818pt}{0.400pt}}
\put(181.0,372.0){\rule[-0.200pt]{4.818pt}{0.400pt}}
\put(161,372){\makebox(0,0)[r]{30}}
\put(1460.0,372.0){\rule[-0.200pt]{4.818pt}{0.400pt}}
\put(181.0,441.0){\rule[-0.200pt]{4.818pt}{0.400pt}}
\put(161,441){\makebox(0,0)[r]{40}}
\put(1460.0,441.0){\rule[-0.200pt]{4.818pt}{0.400pt}}
\put(181.0,511.0){\rule[-0.200pt]{4.818pt}{0.400pt}}
\put(161,511){\makebox(0,0)[r]{50}}
\put(1460.0,511.0){\rule[-0.200pt]{4.818pt}{0.400pt}}
\put(181.0,581.0){\rule[-0.200pt]{4.818pt}{0.400pt}}
\put(161,581){\makebox(0,0)[r]{60}}
\put(1460.0,581.0){\rule[-0.200pt]{4.818pt}{0.400pt}}
\put(181.0,650.0){\rule[-0.200pt]{4.818pt}{0.400pt}}
\put(161,650){\makebox(0,0)[r]{70}}
\put(1460.0,650.0){\rule[-0.200pt]{4.818pt}{0.400pt}}
\put(181.0,720.0){\rule[-0.200pt]{4.818pt}{0.400pt}}
\put(161,720){\makebox(0,0)[r]{80}}
\put(1460.0,720.0){\rule[-0.200pt]{4.818pt}{0.400pt}}
\put(181.0,789.0){\rule[-0.200pt]{4.818pt}{0.400pt}}
\put(161,789){\makebox(0,0)[r]{90}}
\put(1460.0,789.0){\rule[-0.200pt]{4.818pt}{0.400pt}}
\put(181.0,859.0){\rule[-0.200pt]{4.818pt}{0.400pt}}
\put(161,859){\makebox(0,0)[r]{100}}
\put(1460.0,859.0){\rule[-0.200pt]{4.818pt}{0.400pt}}
\put(336.0,163.0){\rule[-0.200pt]{0.400pt}{4.818pt}}
\put(336,122){\makebox(0,0){20}}
\put(336.0,839.0){\rule[-0.200pt]{0.400pt}{4.818pt}}
\put(500.0,163.0){\rule[-0.200pt]{0.400pt}{4.818pt}}
\put(500,122){\makebox(0,0){40}}
\put(500.0,839.0){\rule[-0.200pt]{0.400pt}{4.818pt}}
\put(663.0,163.0){\rule[-0.200pt]{0.400pt}{4.818pt}}
\put(663,122){\makebox(0,0){60}}
\put(663.0,839.0){\rule[-0.200pt]{0.400pt}{4.818pt}}
\put(826.0,163.0){\rule[-0.200pt]{0.400pt}{4.818pt}}
\put(826,122){\makebox(0,0){80}}
\put(826.0,839.0){\rule[-0.200pt]{0.400pt}{4.818pt}}
\put(990.0,163.0){\rule[-0.200pt]{0.400pt}{4.818pt}}
\put(990,122){\makebox(0,0){100}}
\put(990.0,839.0){\rule[-0.200pt]{0.400pt}{4.818pt}}
\put(1153.0,163.0){\rule[-0.200pt]{0.400pt}{4.818pt}}
\put(1153,122){\makebox(0,0){120}}
\put(1153.0,839.0){\rule[-0.200pt]{0.400pt}{4.818pt}}
\put(1317.0,163.0){\rule[-0.200pt]{0.400pt}{4.818pt}}
\put(1317,122){\makebox(0,0){140}}
\put(1317.0,839.0){\rule[-0.200pt]{0.400pt}{4.818pt}}
\put(1480.0,163.0){\rule[-0.200pt]{0.400pt}{4.818pt}}
\put(1480,122){\makebox(0,0){160}}
\put(1480.0,839.0){\rule[-0.200pt]{0.400pt}{4.818pt}}
\put(181.0,163.0){\rule[-0.200pt]{312.929pt}{0.400pt}}
\put(1480.0,163.0){\rule[-0.200pt]{0.400pt}{167.666pt}}
\put(181.0,859.0){\rule[-0.200pt]{312.929pt}{0.400pt}}
\put(41,511){\makebox(0,0){$\epsilon$}}
\put(830,61){\makebox(0,0){$N$}}
\put(181.0,163.0){\rule[-0.200pt]{0.400pt}{167.666pt}}
\put(181,859){\usebox{\plotpoint}}
\multiput(181.59,810.43)(0.488,-15.286){13}{\rule{0.117pt}{11.700pt}}
\multiput(180.17,834.72)(8.000,-207.716){2}{\rule{0.400pt}{5.850pt}}
\put(189.0,627.0){\rule[-0.200pt]{311.002pt}{0.400pt}}
\put(181,859){\usebox{\plotpoint}}
\multiput(181,859)(0.715,-20.743){12}{\usebox{\plotpoint}}
\multiput(189,627)(1.968,-20.662){4}{\usebox{\plotpoint}}
\multiput(197,543)(2.610,-20.591){3}{\usebox{\plotpoint}}
\multiput(206,472)(3.975,-20.371){2}{\usebox{\plotpoint}}
\multiput(214,431)(4.503,-20.261){2}{\usebox{\plotpoint}}
\put(225.24,384.87){\usebox{\plotpoint}}
\put(231.83,365.19){\usebox{\plotpoint}}
\put(239.46,345.90){\usebox{\plotpoint}}
\put(248.92,327.46){\usebox{\plotpoint}}
\put(260.27,310.09){\usebox{\plotpoint}}
\put(273.02,293.73){\usebox{\plotpoint}}
\put(287.33,278.71){\usebox{\plotpoint}}
\put(303.79,266.14){\usebox{\plotpoint}}
\put(321.38,255.13){\usebox{\plotpoint}}
\put(339.76,245.59){\usebox{\plotpoint}}
\put(358.97,237.76){\usebox{\plotpoint}}
\put(378.68,231.37){\usebox{\plotpoint}}
\put(398.62,225.75){\usebox{\plotpoint}}
\put(418.78,220.81){\usebox{\plotpoint}}
\put(439.09,216.73){\usebox{\plotpoint}}
\put(459.45,212.94){\usebox{\plotpoint}}
\put(479.94,209.77){\usebox{\plotpoint}}
\put(500.48,206.94){\usebox{\plotpoint}}
\put(521.07,204.37){\usebox{\plotpoint}}
\put(541.68,202.00){\usebox{\plotpoint}}
\put(562.33,200.33){\usebox{\plotpoint}}
\put(582.99,198.75){\usebox{\plotpoint}}
\put(603.64,197.00){\usebox{\plotpoint}}
\put(624.32,195.71){\usebox{\plotpoint}}
\put(644.97,194.00){\usebox{\plotpoint}}
\put(665.65,192.67){\usebox{\plotpoint}}
\put(686.36,192.00){\usebox{\plotpoint}}
\put(707.03,190.62){\usebox{\plotpoint}}
\put(727.75,190.00){\usebox{\plotpoint}}
\put(748.42,188.57){\usebox{\plotpoint}}
\put(769.14,187.98){\usebox{\plotpoint}}
\put(789.83,187.00){\usebox{\plotpoint}}
\put(810.53,186.00){\usebox{\plotpoint}}
\put(831.25,185.42){\usebox{\plotpoint}}
\put(851.98,185.00){\usebox{\plotpoint}}
\put(872.67,184.00){\usebox{\plotpoint}}
\put(893.42,183.82){\usebox{\plotpoint}}
\put(914.12,183.00){\usebox{\plotpoint}}
\put(934.86,182.77){\usebox{\plotpoint}}
\put(955.57,182.00){\usebox{\plotpoint}}
\put(976.31,181.63){\usebox{\plotpoint}}
\put(997.03,181.00){\usebox{\plotpoint}}
\put(1017.78,181.00){\usebox{\plotpoint}}
\put(1038.48,180.00){\usebox{\plotpoint}}
\put(1059.24,180.00){\usebox{\plotpoint}}
\put(1079.93,179.01){\usebox{\plotpoint}}
\put(1100.69,179.00){\usebox{\plotpoint}}
\put(1121.44,179.00){\usebox{\plotpoint}}
\put(1142.14,178.00){\usebox{\plotpoint}}
\put(1162.89,178.00){\usebox{\plotpoint}}
\put(1183.65,178.00){\usebox{\plotpoint}}
\put(1204.34,177.00){\usebox{\plotpoint}}
\put(1225.10,177.00){\usebox{\plotpoint}}
\put(1245.85,177.00){\usebox{\plotpoint}}
\put(1266.61,177.00){\usebox{\plotpoint}}
\put(1287.30,176.00){\usebox{\plotpoint}}
\put(1308.06,176.00){\usebox{\plotpoint}}
\put(1328.81,176.00){\usebox{\plotpoint}}
\put(1349.57,176.00){\usebox{\plotpoint}}
\put(1370.29,175.46){\usebox{\plotpoint}}
\put(1391.02,175.00){\usebox{\plotpoint}}
\put(1411.77,175.00){\usebox{\plotpoint}}
\put(1432.53,175.00){\usebox{\plotpoint}}
\put(1453.28,175.00){\usebox{\plotpoint}}
\put(1474.02,174.75){\usebox{\plotpoint}}
\put(1480,174){\usebox{\plotpoint}}
\end{picture}
\caption{Comparison of efficiencies when $\mathcal P$ is the
identity permutation and in the case of permutation (\ref{pipe}),
for $1\leq N\leq160$.}
\label{cmp12-efficiency}
\end{figure}
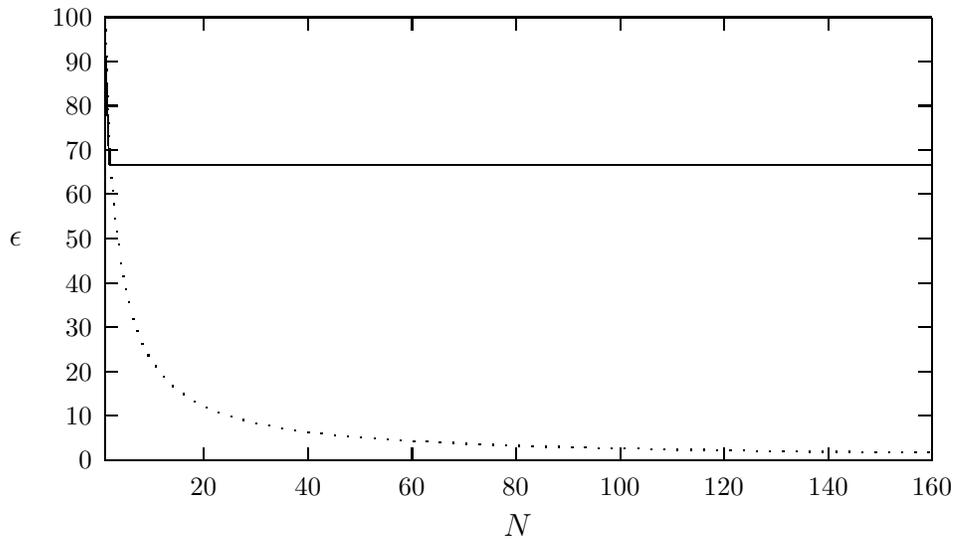

It is possible to show that the distributed algorithm
described in Fig.~\ref{fsa}, with $\mathcal P$ as in
Eq.~(\ref{pipe}), can be computed in linear time:

\begin{prop}\label{prop2}
\[  \lambda_N = 3N. \]
\end{prop}
\begin{pf}

Let us consider run-table $N+1$. Let us strip off its last row; then remove 
each occurrence of $S_{N+1}$, shifting each row leftwards of one 
position.
Remove also each occurrence of $R_{N+1}$. Finally, remove the last
column, now empty because of the previous rules.

Our first goal is showing that what remains is run-table $N$.
To this end, let us remind the reader that each occurrence of an $S_{N+1}$ action only
affects row $N+1$, which has been cut out. Furthermore, each occurrence of $R_{N+1}$ comes from
an $S$ action in row $N+1$. Finally, due to the structure of the permutation,
the last action in row $N+1$ has to be an $S_N$---as a consequence, row $N$ shall contain
an $R_{N+1}$, and remnant rows shall contain action ``$-$''. Removing the $R_{N+1}$ allows
to remove the last column as well, with no coherency violation and no redundant steps.
This proves our first claim.

With a reasoning similar to the one followed for Prop.~\ref{prop1}
we have that
\begin{equation}\label{left2}
\sigma(k+1)= \bigl( \sigma(k) + 3(k+1) \bigr) \frac{k+2}{k+1}.
\end{equation}

that is, by Definition~\ref{sigma},
\begin{equation}\label{recurel2}
\lambda(k+1) = \lambda(k) + 3.
\end{equation}

Recursive relation~(\ref{recurel2}) represents
the first (or forward) difference of $\lambda(k)$ (see e.g., \cite{PaWi79}).
The solution of the above is $\lambda(k) = 3k$.\qed
\end{pf}

The efficiency of the algorithm of pipelined gossiping does not depend
on $N$:
\begin{prop}
$\forall k>0: \varepsilon_k = 2/3$.
\end{prop}
\begin{pf}
Again, let $U(k)$ be the number of used slots in a run of $k$ processors.
From Prop.~\ref{prop2} we know that run-table $k$ differs from run-table $k+1$
only for $k+1$ ``$S$'' actions, $k+1$ ``$R$'' actions, and the last row
consisting of another $k+1$ pairs of useful actions plus some 
non-useful actions.  We conclude that 
\begin{equation}\label{Urecrel}
U(k+1) = U(k)+4(k+1).
\end{equation}

Via e.g., the method of trial solutions for constant coefficient difference
equations introduced in~\cite[p.~16]{PaWi79}, we get to
$U(k)=2k(k+1)$ which obviously satisfies recursive relation~(\ref{Urecrel})
being $2k(k+1)+4(k+1)=2(k+1)(k+2)$.

So
\begin{equation}
\varepsilon_k=\frac{U(k)}{\sigma(k)}=\frac{2k(k+1)}{\lambda_k(k+1)}=\frac23.
\end{equation}
\qed
\end{pf}

\begin{prop}
$\forall k>0: \mu_k = \frac23 (k+1)$.
\end{prop}
\begin{pf}
The proof follows immediately from 
\[ \mu_k = U(k)/\lambda_k = 2k(k+1)/(3k).\]
\qed
\end{pf}

Table~\ref{pipeline} shows how a run-table looks like when multiple
gossiping sessions take place one after the other.
As a result, the central area corresponding to the best observable
performance is prolonged. In such an area, $\varepsilon$ has been
experimentally found to be equal to $N/(N+1)$ and the throughput,
or the number of fully completed gossipings per time step, has been
found to be equal to $t/2$, $t$ being the duration of a time step.
In other words, within that area a gossiping is fully completed every
two time steps in the average. A number of algorithms which are based
on multiple gossipings may greatly benefit from this approach, e.g.,
those implemented in the distributed voting algorithm described 
in~\cite{DeDL98e}.

\begin{table}
\begin{small}
\hspace{-30pt}
\begin{tabular}{l|c@{\hspace{1pt}}%
c@{\hspace{1pt}}c@{\hspace{1pt}}c@{\hspace{1pt}}c@{\hspace{1pt}}c@{\hspace{1pt}}c@{\hspace{1pt}}c@{\hspace{1pt}}c@{\hspace{1pt}}%
    c@{\hspace{1pt}}c@{\hspace{1pt}}c@{\hspace{1pt}}c@{\hspace{1pt}}c@{\hspace{1pt}}%
    c@{\hspace{1pt}}c@{\hspace{1pt}}c@{\hspace{1pt}}c@{\hspace{1pt}}c@{\hspace{1pt}}%
    c@{\hspace{1pt}}%
%
    c@{\hspace{1pt}}c@{\hspace{1pt}}c@{\hspace{1pt}}c@{\hspace{1pt}}c@{\hspace{1pt}}%
    c@{\hspace{1pt}}c@{\hspace{1pt}}c@{\hspace{1pt}}c@{\hspace{1pt}}c@{\hspace{1pt}}%
c@{\hspace{1pt}}c@{\hspace{1pt}}c@{\hspace{1pt}}c@{\hspace{1pt}}c@{\hspace{1pt}}c@{\hspace{1pt}}c@{\hspace{1pt}}c@{\hspace{1pt}}}
&\tiny1&\tiny2&\tiny3&%
\tiny4&\tiny5&\tiny6&\tiny7&\tiny8&\tiny9&\tiny10&\tiny11&\tiny12&\tiny13&\tiny14&\tiny15&\tiny16%
&\tiny17&\tiny18&\tiny19%
&$\dots$%
\\ \hline
\sf 0&$S_{1}$&$S_{2}$&$S_{3}$&$S_{4}$&$-$&$R_{1}$&$R_{2}$&$R_{3}$&$R_{4}$%
       &$\curvearrowright$&$S_1$&$S_2$&$S_{3}$&$S_{4}$&$-$&$R_{1}$&$R_{2}$&$R_{3}$&$R_{4}$%
	&$\dots$%
       &$\curvearrowright$&$S_1$&$S_2$&$S_{3}$&$S_{4}$&$-$&$R_{1}$&$R_{2}$&$R_{3}$&$R_{4}$%
&$-$&$-$&$-$\\
\sf 1&$R_{0}$&$\curvearrowright$&$S_{2}$&$S_{3}$&$S_{4}$&$S_{0}$&$-$&$R_{2}$&$R_{3}$%
       &$R_{4}$&$R_0$&$\curvearrowright$&$S_{2}$&$S_{3}$&$S_{4}$&$S_{0}$&$-$&$R_{2}$&$R_{3}$%
	&$\dots$%
       &$R_{4}$&$R_0$&$\curvearrowright$&$S_{2}$&$S_{3}$&$S_{4}$&$S_{0}$&$-$&$R_{2}$&$R_{3}$%
&$R_{4}$&$-$&$-$\\
\sf 2&$-$&$R_{0}$&$R_{1}$&$\curvearrowright$&$S_{3}$&$S_{4}$&$S_{0}$&$S_{1}$&$-$%
       &$R_{3}$&$R_{4}$&$R_0$&$R_{1}$&$\curvearrowright$&$S_{3}$&$S_{4}$&$S_{0}$&$S_{1}$&$-$%
	&$\dots$%
       &$R_{3}$&$R_{4}$&$R_0$&$R_{1}$&$\curvearrowright$&$S_{3}$&$S_{4}$&$S_{0}$&$S_{1}$&$-$%
&$R_{3}$&$R_{4}$&$-$\\
\sf 3&$-$&$-$&$R_{0}$&$R_{1}$&$R_{2}$&$\curvearrowright$&$S_{4}$&$S_{0}$&$S_{1}$%
       &$S_{2}$&$-$&$R_{4}$&$R_{0}$&$R_{1}$&$R_{2}$&$\curvearrowright$&$S_{4}$&$S_{0}$&$S_{1}$%
	&$\dots$%
       &$S_{2}$&$-$&$R_{4}$&$R_{0}$&$R_{1}$&$R_{2}$&$\curvearrowright$&$S_{4}$&$S_{0}$&$S_{1}$%
&$S_{2}$&$-$&$R_{4}$\\
\sf 4&$-$&$-$&$-$&$R_{0}$&$R_{1}$&$R_{2}$&$R_{3}$&$\curvearrowright$&$S_{0}$%
       &$S_{1}$&$S_{2}$&$S_{3}$&$-$&$R_{0}$&$R_{1}$&$R_{2}$&$R_{3}$&$\curvearrowright$&$S_{0}$%
	&$\dots$%
       &$S_{1}$&$S_{2}$&$S_{3}$&$-$&$R_{0}$&$R_{1}$&$R_{2}$&$R_{3}$&$\curvearrowright$&$S_{0}$%
&$S_{1}$&$S_{2}$&$S_{3}$\\
\hline
&2&2&4&4&4&4&4&4&4&4&4&4&4&4&4&4&4&4&4%
	&$\dots$%
&4&4&4&4&4&4&4&4&4&4&4&2&2
\end{tabular}
\end{small}
\caption{The algorithm is modified so that multiple gossiping sessions take place.
The central, best performing area is consequently prolonged. Therein $\varepsilon$ is 
equal to $N/(N+1)$. 
Note how within that area there are consecutive ``zones'' of ten columns each,
within whom five gossiping sessions reach their conclusion. For instance,
such a zone is the region between columns 7 and 16: therein, at entries $(4,7)$,
$(0,9)$, $(1,10)$, $(2,11)$, and $(3,12)$, a processor gets the last value
of a broadcast and can perform some work on a full set of values. This
brings to a throughput of $t/2$, where $t$ is the duration of a slot.}
\label{pipeline}
\end{table}

Obviously our model reaches such a performance only if the system
calls for exactly one time step to communicate between any two processors,
like e.g. in a crossbar system. This is similar to the constraint
of hardware pipelined processors which call for a number of
memory ports equal to $n$, $n$ being the number of pipeline 
stages supported by that machine---in this way
the system is able to overlap any two of its stages. This of course
turns into requiring to have a memory system 
capable of delivering $n$ times the original bandwidth
of a corresponding, unpipelined machine~\cite{PaHe96}.

\begin{figure}
\centerline{\psfig{figure=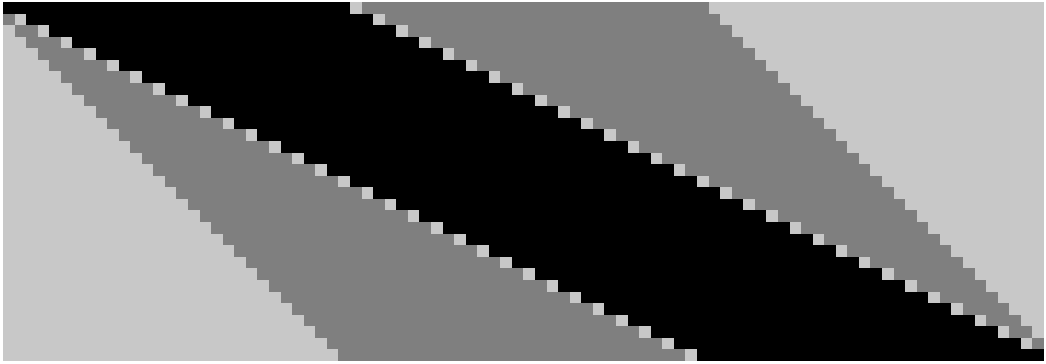,height=\textwidth,angle=-90}}
\caption{A graphical representation for run-table 30 when $\mathcal P$ is
permutation~(\ref{pipe}).}\label{s30}
\end{figure}

Note how the specularity of graphs like the one of Fig.~\ref{s30} translates
into a palindrome $\vec\nu$ string.


\section{Further Optimizations}\label{opt}
Evidently the execution of ``$-$'' and ``$\curvearrowright$'' actions---the
``bubbles''---is an impairment towards the optimum.
As a consequence, a general strategy to increase the
performance of our algorithms could be the following one:
\begin{enumerate}
\item execute a base rule (corresponding to the adoption of any
permutation $\mathcal P$, like e.g., those presented in \S\ref{zerocase},
\S\ref{randcase}, or \S\ref{tao-pb}), and
\item as soon as there is a wait-in-sending, choose a different
destination between those that would execute a wait-in-receiving.
\end{enumerate}

In other words, the processor who gains the right to broadcast
does follow the order given by $\mathcal P$ unless it knows
(by calculating its own run-table) that doing that it would trigger
a wait-to-send action. In such latter case, it looks for another
candidate among those following the current one in permutation $\mathcal P$.
If there exists at least one such processor that would otherwise
be wasting its slot in a wait-in-receiving, then the message
is sent to it. In some sense, this allows each 
broadcasting processor to rearrange its leading $\mathcal P$
into a ``better'' $\mathcal{P}'$, driven by the possibility
to (locally) increase the number of used slots. Of course
this local gain might turn into a loss later on. In this
Section we describe how the above procedure perturbs
the values of $\lambda$, $\mu$, and $\varepsilon$ for the
so far discussed three cases.

This strategy depicts similarities with the optimization method
known as Pipeline Scheduling and described in~\cite{PaHe96}:
for any program $p$ whose run corresponds to the execution of $\mathrm{IC}(p)$ instructions
($\mathrm{IC}(p)$ = instruction count of program $p$), namely
\begin{equation}\label{Ik}
(I_k)_{1\leq k\leq \mathrm{IC}(p)},
\end{equation}

the detection of an obstacle to the optimum (a \emph{stall\/})
triggers an attempt to go round it by trying to rearrange (\ref{Ik}) as
\begin{equation}\label{I'j}
(I'_j)_{1\leq j\leq \mathrm{IC}(p)},
\end{equation}

where (\ref{I'j}) is substantially a (semantically equivalent)
permutation of (\ref{Ik}) such that the obstacle is removed.

Of course in our case semantical equivalence is guaranteed by
the fact that each processor just modifies its permutation on-the-fly. This
does not affect the output of the algorithm, so our task is much easier
than if we had to implement actual pipeline scheduling.

The following algorithm can be used to simulate a run and compute
the entire broadcasting sequence i.e.,
the second row of Fig.~\ref{fsa}:

\begin{alg}{\emph{: Gossiping with permutation scheduling}}
\begin{tabbing}
xx\=xx\=xx\=xx\=xx\=xx\=xx\=xx\=xx\=xx\=xx\=xx\=xxxxx\=xx\kill
         \> \> \emph{Input:\/} $N$, $\mathcal{P}$, $p$ (processor id)\\
         \> \> \emph{Input:\/} $\mathrm{run}$ (running run-table), $t$ (current time step)\\
         \> \> \emph{Input:\/} $m$ (message to be broadcast)\\
         \> \> \emph{Output:\/} $\mathrm{run}$, $t$\\
{\bf 1} \> \> \keyw{begin}\\
{\bf 2} \> \> \> $\vec{f} := \mathrm{TRUE}$ \>\>\>\>\>\>\>\>\>\> \{ set each entry of $\vec{f}$ to $\mathrm{TRUE}$ \} \\
{\bf 3} \> \> \> $i := 1$ \\
{\bf 4} \> \> \> $w := 0$ \\
{\bf 5} \> \> \> \keyw{while} $i\leq N$ \keyw{do} \>\>\>\>\>\>\>\>\>\> \{ for each symbol of $\mathcal P$ \} \\
{\bf 6} \> \> \> \> $j := i + t + w -1$ \\

         \> \> \> \> \{ if $\mathcal{P}_i$ has never been used and processor $i$ is available \} \\
{\bf 7} \> \> \> \> \keyw{if} $f_i = \mathrm{TRUE}  \ \wedge \ \mathrm{run}(\mathcal{P}_i, j) = \mathrm{FREE}$ 
                     \keyw{then} \\

{\bf 8} \> \> \> \> \> $f_i := \mathrm{FALSE}$ \>\>\>\>\>\>\>\> \{ mark $\mathcal{P}_i$ as used \} \\
{\bf 9} \> \> \> \> \> Send $m$ to processor $\mathcal{P}_i$ \\
{\bf 10} \> \> \> \> \> $\mathrm{run}(p,  j) :=  p\, S^{j}\, \mathcal{P}_i$ \\
{\bf 11} \> \> \> \> \> $\mathrm{run}(\mathcal{P}_i, j) :=  \mathcal{P}_i\, R^{j}\, p$ \\
{\bf 12} \> \> \> \> \> $i := i + 1$ \>\>\>\>\>\>\>\> \{ go to next item of $\mathcal{P}$ \} \\
         \> \> \> \> \{ if $\mathcal{P}_i$ has been already used or processor $i$ is not available \} \\
{\bf 13} \> \> \> \> \keyw{else} \>\>\>\>\>\>\>\>\> \{ i.e., when $f_i = \mathrm{FALSE} \ \vee \ \mathrm{run}(\mathcal{P}_i, j) \neq \mathrm{FREE}$ \} \\
         \> \> \> \> \> \{ orderly search for a possible substitute \} \\
{\bf 14} \> \> \> \> \> $\mathrm{stop} := \mathrm{FALSE}$ \\
{\bf 15} \> \> \> \> \> $l := 1$ \\
{\bf 16} \> \> \> \> \> \keyw{while} $l \leq N \ \wedge \ \mathrm{stop} = \mathrm{FALSE}$ \keyw{do} \\
{\bf 17} \> \> \> \> \> \> \keyw{if} $f_l = \mathrm{TRUE}  \ \wedge \ \mathrm{run}(\mathcal{P}_l, j) = \mathrm{FREE}$
			   \keyw{then}\\
{\bf 18} \> \> \> \> \> \> \> $\mathrm{stop} := \mathrm{TRUE}$\\
{\bf 19} \> \> \> \> \> \> \keyw{else}\\
{\bf 20} \> \> \> \> \> \> \> $l := l+1$\\
{\bf 21} \> \> \> \> \> \> \keyw{endif}\\
{\bf 22} \> \> \> \> \> \keyw{enddo}\\
         \> \> \> \> \> \{ if a candidate has been found at entry $l$, use $\mathcal{P}_l$ instead of $\mathcal{P}_i$ \} \\
{\bf 23} \> \> \> \> \> \keyw{if} $\mathrm{stop} = \mathrm{TRUE}$ \keyw{then} \\
{\bf 24} \> \> \> \> \> \> $f_l := \mathrm{FALSE}$ \\
{\bf 25} \> \> \> \> \> \> Send $m$ to processor $\mathcal{P}_l$ \\
{\bf 26} \> \> \> \> \> \> $\mathrm{run}(p,  j) :=  p\, S^{j}\, \mathcal{P}_l$ \\
{\bf 27} \> \> \> \> \> \> $\mathrm{run}(\mathcal{P}_l, j) :=  \mathcal{P}_l\, R^{j}\, p$ \\
{\bf 28} \> \> \> \> \> \> $i := i + 1$ \\
{\bf 29} \> \> \> \> \> \keyw{else} \>\>\>\>\>\>\>\> \{ if no such an $l$ exists\ldots \} \\

{\bf 30} \> \> \> \> \> \> $\mathrm{run}(p,  j) :=  \ \curvearrowright$ \>\>\>\>\>\>\> \{ store a wait-for-sending \} \\
         \> \> \> \> \> \> \{ deal again with the current value of $i$, but on next column \}\\
{\bf 31} \> \> \> \> \> \> $w := w+1$ \\
{\bf 32} \> \> \> \> \> \keyw{endif}\\
{\bf 33} \> \> \> \> \keyw{endif}\\
{\bf 34} \> \> \> \keyw{enddo}\\
{\bf 35} \> \> \> $t := t + N + w$\\
{\bf 36} \> \> \keyw{end}.
\end{tabbing}
\label{algo}
\end{alg}

\subsection{Applying Algorithm~\ref{algo} to the 
	    Case of the Identity Permutation}\label{algoz+a}

\begin{figure}
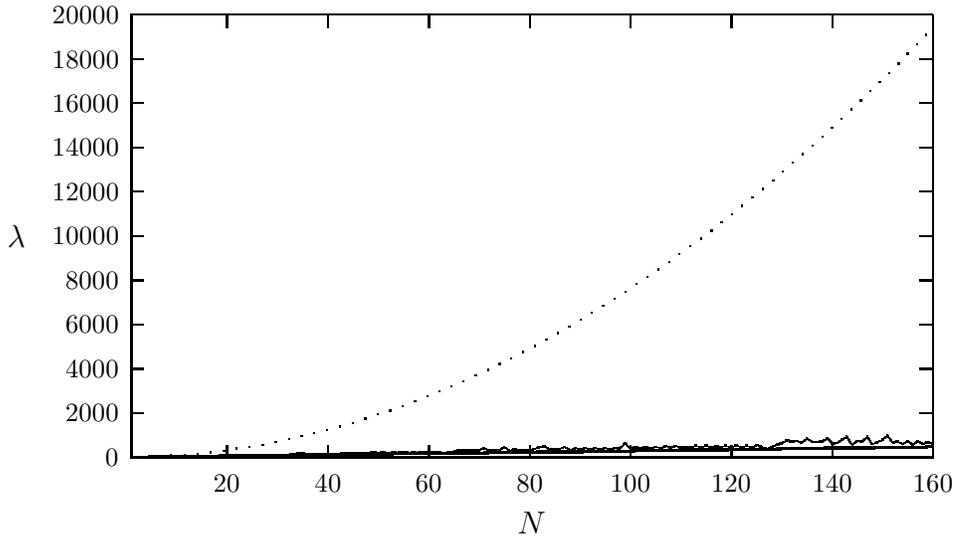

\input compare-steps.z+a
\caption{This picture portrays and compares run lengths for $1\leq N\leq160$ when
$\mathcal P$ is the identity permutation (dotted parabola), when Algorithm~\ref{algo}
is applied to the case of the identity permutation (piecewise line), and in the case
of the pipelined gossiping.
Note how Algorithm~\ref{algo} always improves its base
method, and in a small number of cases ($N=2^i-1, i\leq10$) it reaches a better performance than
that of pipelined gossiping (see Table~\ref{pots}). This is also shown in
Fig.~\ref{compare-efficiency.z+a}.}
\label{compare-steps.z+a}
\end{figure}

\begin{figure}
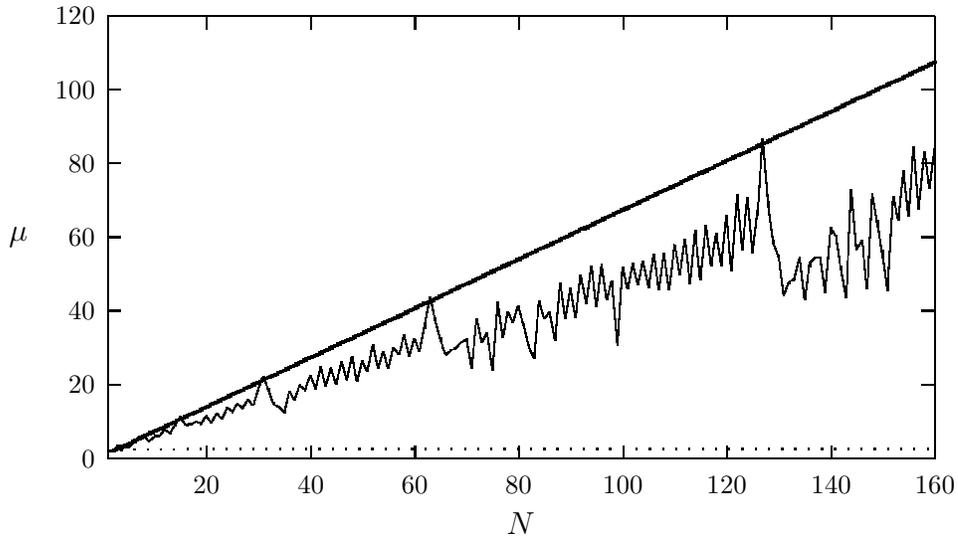

\input compare-average.z+a
\caption{Values of $\mu$ for the three cases of Fig.~\ref{compare-steps.z+a}.}
\label{compare-average.z+a}
\end{figure}

\begin{figure}
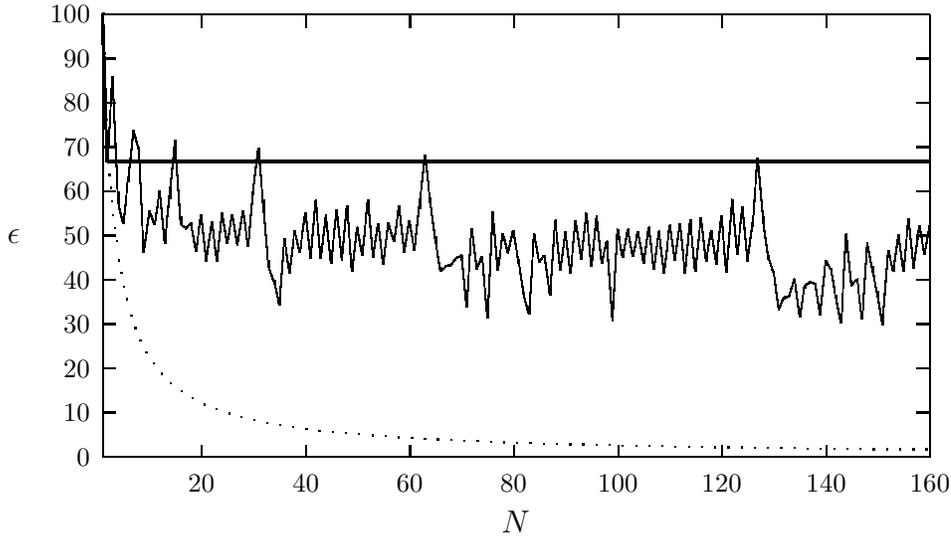

\input compare-efficiency.z+a
\caption{Values of $\varepsilon$ for the three cases of Fig.~\ref{compare-steps.z+a}.
Peek values are in Table~\ref{pots}.}
\label{compare-efficiency.z+a}
\end{figure}

Figures~\ref{compare-steps.z+a}, \ref{compare-average.z+a}, and
\ref{compare-efficiency.z+a} describe the improvement we observed
by applying optimizing Algorithm~\ref{algo} to the case
of the identity permutation.

We experimentally found that the strategy does greatly improve the values of
$\mu$, $\lambda$, and $\varepsilon$.
We also observed that a particularly good case occurs
when the number of processors employed is a power of two. Table~\ref{rt7}
for instance shows run-table 7, that has an efficiency of
$73.68\%$. This efficiency though tends to decrease. In particular
we found that, after the case of $N=2^{10}-1$, the efficiency 
becomes lower than $2/3$ i.e., the one of the algorithm
of pipelined gossiping (see Table~\ref{pots}.)

\begin{table}
\begin{center}
\begin{tabular}{l|c@{\hspace{1pt}}c@{\hspace{1pt}}c@{\hspace{1pt}}c@{\hspace{1pt}}c@{\hspace{1pt}}c@{\hspace{1pt}}c@{\hspace{1pt}}c@{\hspace{1pt}}c@{\hspace{1pt}}c@{\hspace{1pt}}c@{\hspace{1pt}}c@{\hspace{1pt}}c@{\hspace{1pt}}c@{\hspace{1pt}}c@{\hspace{1pt}}c@{\hspace{1pt}}c@{\hspace{1pt}}c@{\hspace{1pt}}c@{\hspace{1pt}}}
$\stackrel{\hbox{\sffamily\scriptsize id}}{\downarrow}$ $\stackrel{\hbox{\sffamily\scriptsize step}}{\rightarrow}$&\tiny1&\tiny2&\tiny3&\tiny4&\tiny5&\tiny6&\tiny7&\tiny8&\tiny9&\tiny10&\tiny11&\tiny12&\tiny13&\tiny14&\tiny15&\tiny16&\tiny17&\tiny18&\tiny
19\\ \hline
\sf 0&$S_{1}$&$S_{2}$&$S_{3}$&$S_{4}$&$S_{5}$&$S_{6}$&$S_{7}$&$R_{1}$&$R_{2}$&$R_{3}$&$R_{4}$&$R_{6}$&$R_{5}$&$R_{7}$&$-$&$-$&$-$&$-$&$-$\\
\sf 1&$R_{0}$&$S_{3}$&$S_{2}$&$S_{5}$&$S_{4}$&$S_{7}$&$S_{6}$&$S_{0}$&$R_{3}$&$R_{2}$&$R_{5}$&$R_{4}$&$R_{6}$&$-$&$R_{7}$&$-$&$-$&$-$&$-$\\
\sf 2&$-$&$R_{0}$&$R_{1}$&$S_{3}$&$S_{6}$&$S_{4}$&$S_{5}$&$S_{7}$&$S_{0}$&$S_{1}$&$R_{3}$&$R_{7}$&$R_{4}$&$R_{5}$&$-$&$-$&$-$&$R_{6}$&$-$\\
\sf 3&$-$&$R_{1}$&$R_{0}$&$R_{2}$&$S_{7}$&$S_{5}$&$S_{4}$&$S_{6}$&$S_{1}$&$S_{0}$&$S_{2}$&$R_{5}$&$R_{7}$&$R_{4}$&$R_{6}$&$-$&$-$&$-$&$-$\\
\sf 4&$-$&$-$&$-$&$R_{0}$&$R_{1}$&$R_{2}$&$R_{3}$&$S_{5}$&$S_{6}$&$S_{7}$&$S_{0}$&$S_{1}$&$S_{2}$&$S_{3}$&$R_{5}$&$R_{6}$&$R_{7}$&$-$&$-$\\
\sf 5&$-$&$-$&$-$&$R_{1}$&$R_{0}$&$R_{3}$&$R_{2}$&$R_{4}$&$S_{7}$&$S_{6}$&$S_{1}$&$S_{3}$&$S_{0}$&$S_{2}$&$S_{4}$&$R_{7}$&$R_{6}$&$-$&$-$\\
\sf 6&$-$&$-$&$-$&$-$&$R_{2}$&$R_{0}$&$R_{1}$&$R_{3}$&$R_{4}$&$R_{5}$&$S_{7}$&$S_{0}$&$S_{1}$&$\curvearrowright$&$S_{3}$&$S_{4}$&$S_{5}$&$S_{2}$&$R_{7}$\\
\sf 7&$-$&$-$&$-$&$-$&$R_{3}$&$R_{1}$&$R_{0}$&$R_{2}$&$R_{5}$&$R_{4}$&$R_{6}$&$S_{2}$&$S_{3}$&$S_{0}$&$S_{1}$&$S_{5}$&$S_{4}$&$\curvearrowright$&$S_{6}$\\
\hline
$\stackrel{\hbox{\sffamily\scriptsize $\vec\nu$}}{\rightarrow}$&2&4&4&6&8&8&8&8&8&8&8&8&8&6&6&4&4&2&2
\end{tabular}
\end{center}
\caption{Run-table 7 for $\mathcal P$ equal to the identity permutation,
modified by Algorithm~\ref{algo} ($\mu=5.89$, $\varepsilon=73.68\%$).}
\label{rt7}
\end{table}

\begin{table}
\begin{center}
\begin{tabular}{|c|ccccccccccc|} \hline 
$i$&{1}&{2}&{3}&{4}&{5}&{6}&{7}&{8}&{9}&{10}&{11}\\ \hline
$\varepsilon$&100&85.71&73.68&71.43&69.66&68.11&67.55&67.11&66.88&66.75&65.34\\
\hline
\end{tabular}
\end{center}
\caption{$\varepsilon$ values for different values of $N=2^i-1$.}
\label{pots}
\end{table}


\subsection{Applying Algorithm~\ref{algo} to the Case of the Pseudo-Random Permutation}
\begin{figure}
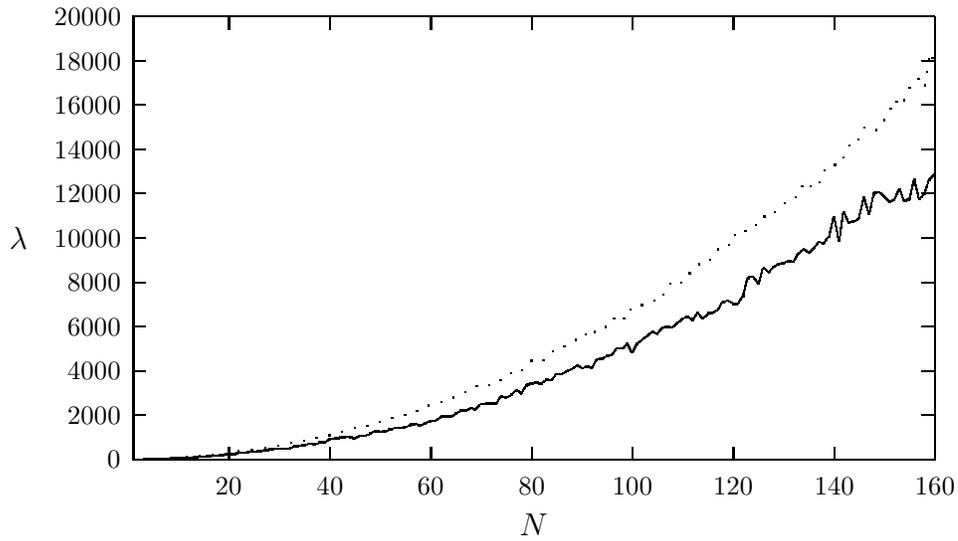

\input compare-steps.r+a
\caption{Comparison of lengths when $\mathcal P$ is a pseudo-random permutation
(dots) and with the addition of Algorithm~\ref{algo} (piecewise line), $1\leq N\leq160$.}
\label{compare-steps.r+a}
\end{figure}

\begin{figure}
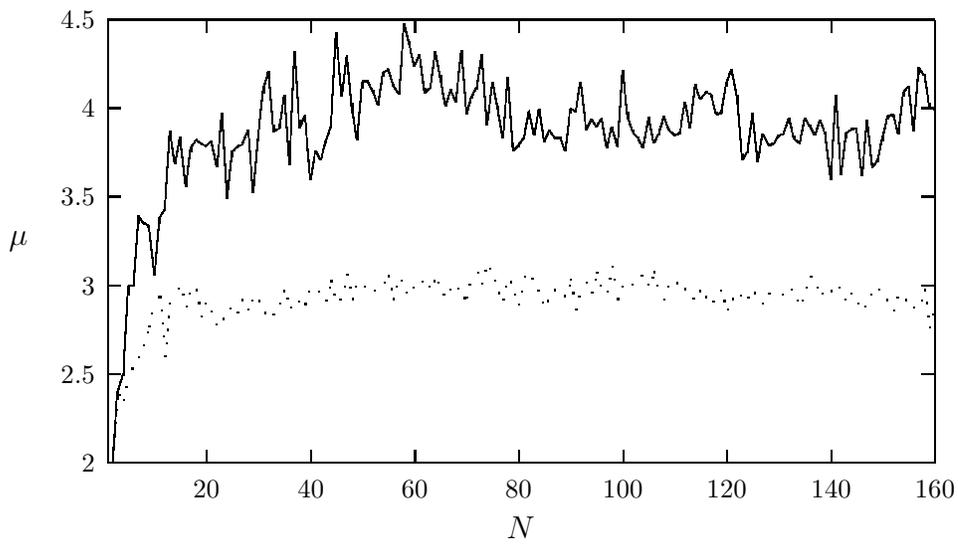

\input compare-average.r+a
\caption{Comparison of the values of $\mu$ in the two cases of Fig.~\ref{compare-steps.r+a}.}
\label{compare-average.r+a}
\end{figure}

\begin{figure}
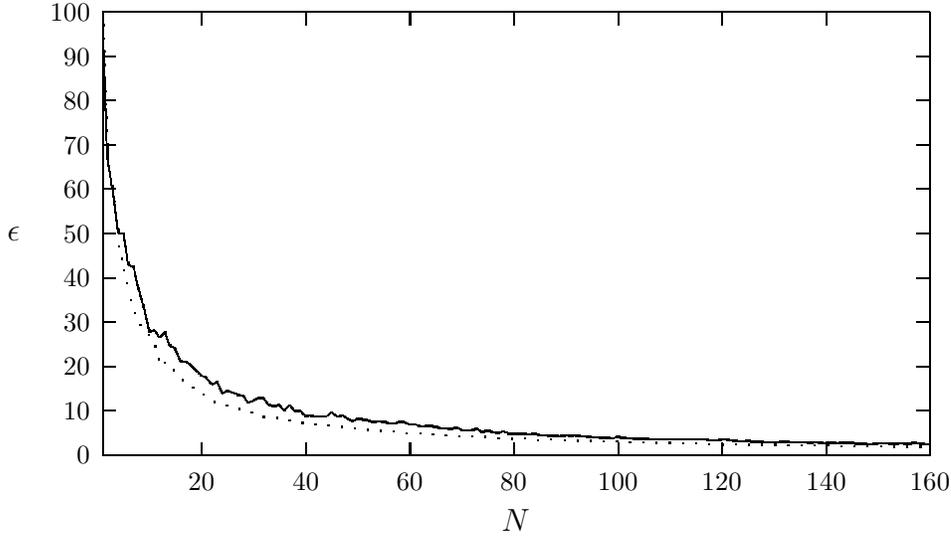

\input compare-efficiency.r+a
\caption{Comparison of the values of $\varepsilon$ in the two cases of Fig.~\ref{compare-steps.r+a}.}
\label{compare-efficiency.r+a}
\end{figure}

Also when applied to the case of the pseudo-random permutation, 
Algorithm~\ref{algo} improves performance---this is shown for $1\leq N\leq160$
in the Fig.~\ref{compare-steps.r+a}, Fig.~\ref{compare-average.r+a}, and Fig.~\ref{compare-efficiency.r+a}.
This time the improvement is not as high as in \S\ref{algoz+a}.

\clearpage

\clearpage

\subsection{Applying Algorithm~\ref{algo} in the Pipelined Broadcast Mode.}
\begin{figure}
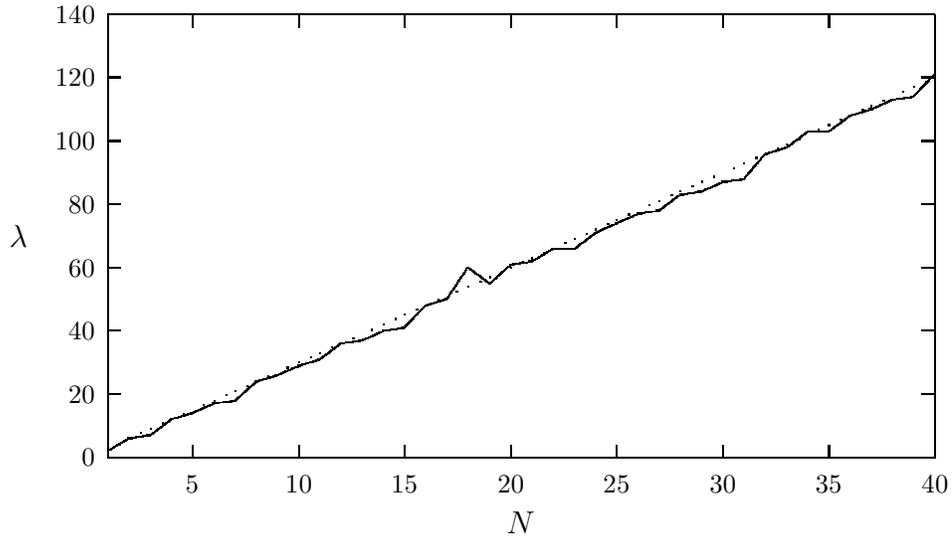

\input compare-steps.s+a
\caption{Values of $\lambda$ for $1\leq N\leq40$ when $\mathcal P$ is (\ref{pipe}),
with (piecewise line) and without (dotted line) the optimization of Algorithm~\ref{algo}.}
\label{compare-steps.s+a}
\end{figure}

\begin{figure}
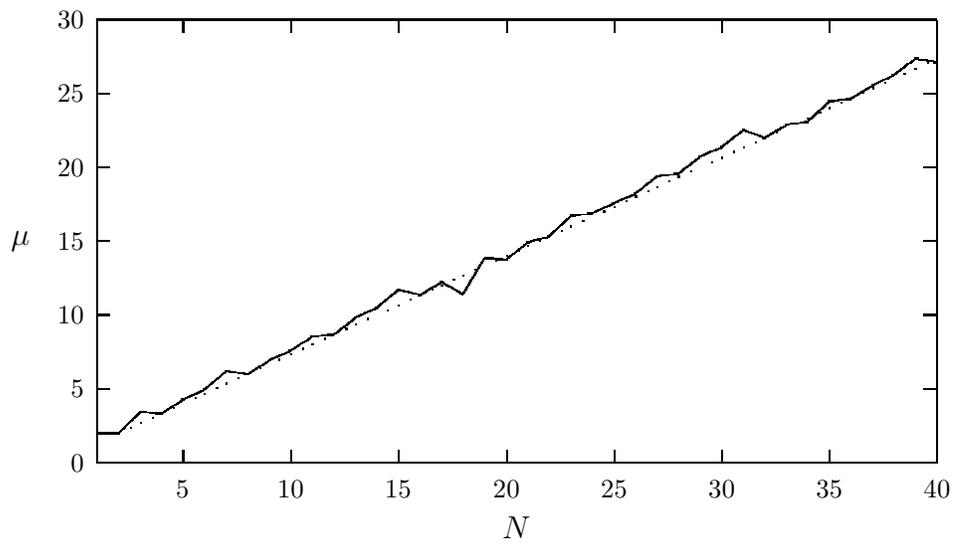

\input compare-average.s+a
\caption{Values of $\mu$ in the two cases of Fig.~\ref{compare-steps.s+a}.}
\label{compare-average.s+a}
\end{figure}

\begin{figure}
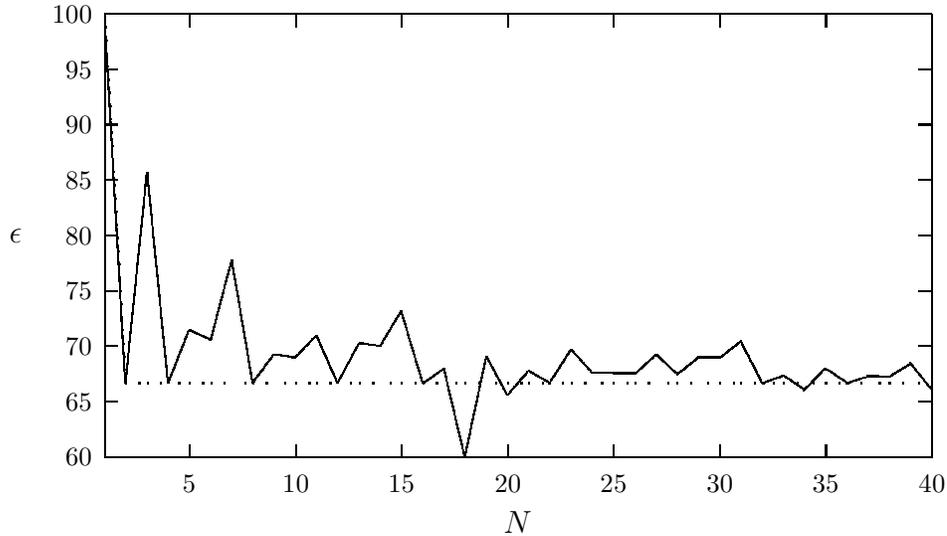

\input compare-efficiency.s+a
\caption{Values of $\varepsilon$ in the two cases of Fig.~\ref{compare-steps.s+a}.}
\label{compare-efficiency.s+a}
\end{figure}

When coupling Algorithm~\ref{algo} to the algorithm of pipelined gossiping,
the local optimizations gave unstable, and in
some cases even negative returns (see Fig.~\ref{compare-steps.s+a}, \ref{compare-average.s+a},
and \ref{compare-efficiency.s+a}). For instance, Table~\ref{taopb+algo} is run-table 4, which 
shows the same values of $\mu$ and $\varepsilon$ as if we had performed no optimization at all.
$N=18$ is an example of negative return---in this case e.g., $\varepsilon$ falls to 60\%.


\begin{table}
\begin{center}
\begin{tabular}{l|c@{\hspace{1pt}}c@{\hspace{1pt}}c@{\hspace{1pt}}c@{\hspace{1pt}}c@{\hspace{1pt}}c@{\hspace{1pt}}c@{\hspace{1pt}}c@{\hspace{1pt}}c@{\hspace{1pt}}c@{\hspace{1pt}}c@{\hspace{1pt}}c@{\hspace{1pt}}}
$\stackrel{\hbox{\sffamily\scriptsize id}}{\downarrow}$ $\stackrel{\hbox{\sffamily\scriptsize step}}{\rightarrow}$&\tiny1&\tiny2&\tiny3&\tiny4&\tiny5&\tiny6&\tiny7&\tiny8&\tiny9&\tiny10&\tiny11&\tiny12\\ \hline
\sf 0&$S_{1}$&$S_{2}$&$S_{3}$&$S_{4}$&$R_{2}$&$R_{1}$&$-$&$R_{3}$&$R_{4}$&$-$&$-$&$-$\\
\sf 1&$R_{0}$&$S_{3}$&$S_{2}$&$\curvearrowright$&$S_{4}$&$S_{0}$&$R_{2}$&$R_{4}$&$R_{3}$&$-$&$-$&$-$\\
\sf 2&$-$&$R_{0}$&$R_{1}$&$S_{3}$&$S_{0}$&$S_{4}$&$S_{1}$&$-$&$-$&$R_{3}$&$R_{4}$&$-$\\
\sf 3&$-$&$R_{1}$&$R_{0}$&$R_{2}$&$\curvearrowright$&$\curvearrowright$&$S_{4}$&$S_{0}$&$S_{1}$&$S_{2}$&$-$&$R_{4}$\\
\sf 4&$-$&$-$&$-$&$R_{0}$&$R_{1}$&$R_{2}$&$R_{3}$&$S_{1}$&$S_{0}$&$\curvearrowright$&$S_{2}$&$S_{3}$\\
\hline
$\stackrel{\hbox{\sffamily\scriptsize $\vec\nu$}}{\rightarrow}$&2&4&4&4&4&4&4&4&4&2&2&2
\end{tabular}
\caption{Run-table 4 in pipelined gossiping mode and applying Algorithm~\ref{algo}.
$\mu=3.33$ slots out of 5, or an efficiency of 66.67\%. In other words, Algorithm~\ref{algo}
affected the run-table without developing any improvement---in particular,
the ending order has changed.}
\label{taopb+algo}
\end{center}
\end{table}

\clearpage

\section{Applicative Examples}\label{vf}
In this section we provide two example applications for the algorithms described
in this paper: a restoring organ (Sect.~\ref{s.vf}) and a proposal for a Hopfield neural network
architecture (Sect.~\ref{s.hop}).

\subsection{The EFTOS Voting Farm}\label{s.vf}
\begin{figure}
\centerline{\psfig{figure=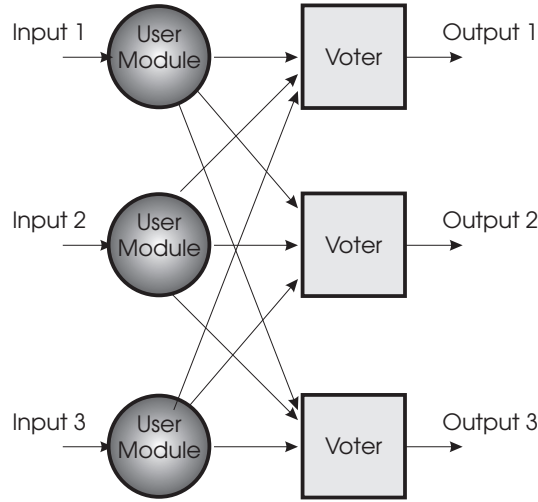,width=7.0cm}}
\caption{A restoring organ~\cite{John89a}, i.e., a $N$\kern-1pt-modular
  redundant system with $N$ voters, when $N=3$. Note that a de-multiplexer
is required to produce the single final output.}
\label{ro}
\end{figure}

\begin{figure}
\centerline{\psfig{figure=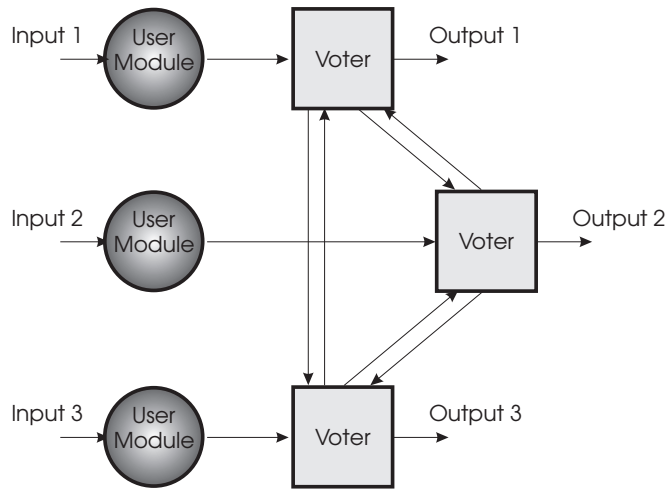,width=8.7cm}}
\caption{Structure of the EFTOS VF for $N=3$.}\label{ronew}
\end{figure}

The EFTOS Voting Farm (VF) is a 
software component that can be used to implement
restoring organs i.e., $N$-modular
redundancy systems (\nmr) with
$N$-replicated voters~\cite{John89a}
(see Fig.~\ref{ro}). Basic design goals of such tools include fault transparency
but also replication transparency,
a high degree of flexibility and ease-of-use, and good performance.
Restoring organs allow to overcome the shortcoming of having
one voter, the failure of which leads to the failure of the whole system
even when each and every other module is still running correctly.
From the point of view of software engineering, such systems
though are characterised by two major drawbacks:
\begin{itemize}
\item Each module in the \nmr{} must be aware of and responsible for
interacting with the whole set of voters;
\item The complexity of these interactions, which is a function that increases
quadratically with $N$ (the cardinality of the set of voters),
burdens each module in the \nmr.
\end{itemize}
To overcome these drawbacks, VF adopts a different procedure, as
described in Fig.~\ref{ronew}: in this new procedure, each module only has
to interact with, and be aware of {\em one\/} voter,
regardless of the value of $N$. 

The VF is an example of an application taking advantage of the algorithms
described in this paper: indeed its voters play the role of the processors 
of Sect.~\ref{model}. In a fully connected and synchronous system
then steady state performance of the VF follows the ones shown in this paper.
In particular this leads to high scalability and performance. 

A thorough description of the VF can be found in~\cite{DeDL98e}.

\subsection{Applications to Hopfield Neural Networks}\label{s.hop}
A well known paradigm of Neural Computing is minimisation~\cite{SFK97}. Such cognitive technique
has been proved to be able to provide satisfactory solutions to two classes of problems:
\begin{enumerate}                       
\item Recognition, where a partial or corrupted pattern is given as input and the action of
      the system network is to recognise it as one of its stored patterns.
\item Discovery of local minima---a typical example being the travelling salesman
      problem~\cite{HT85}.
\end{enumerate}

Hopfield networks~\cite{Hop82,Roj96} have been found to be particularly useful in solving the
above two classes of problems. A Hopfield network substantially is a net of
binary threshold logic units, connected in an all-to-all pattern,
with weighted connections between units. Weights are changed according to the
so-called Hebb rule---that takes over the role of the training step of, e.g.,
the multi-layer perceptron~\cite{Ros58}. Given a partial or corrupted input pattern,
a Hopfield network allows to determine which of the data stored in the network
resemble the most the input pattern. This is achieved by means of an iterative
procedure, the starting point of which is the input pattern, which consists of
serial, element by element updating. This procedure is indeed a
gossiping algorithm. When the number of neurons is large the adoption of
a scalable procedure like the algorithm of pipelined gossiping could provide
a satisfactory solution.
\begin{figure}
\centerline{\psfig{figure=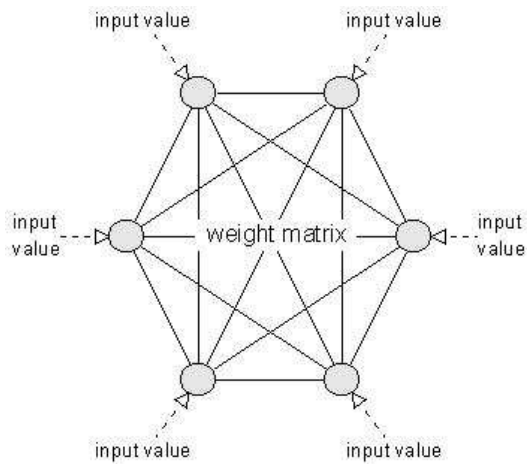,width=0.5\textwidth}}
\caption{A Hopfield Neural Network.}\label{fig:hop}
\end{figure}

\section{Conclusions and Future Work}\label{end}
As e.g., in~\cite{DeFl96}, a formal model for a family of algorithms depending on a
combinatorial parameter, $\mathcal P$, has been introduced and discussed.
Several case studies have been designed, simulated, and analyzed, 
also categorizing in some cases their asymptotic bahaviour.
In one of these cases---the algorithm of pipelined gossiping---it has been proved
that the efficiency of the algorithm does not depend on $N$---a result that
overcomes those of all the known gossiping algorithms~\cite{KruCV92,Hrom96}.
An optimizing algorithm has been presented and discussed as well.

We experimentally found that the efficiencies of the base cases,
improved via the optimization algorithm, lay in general quite ``close'' to
the efficiency of the algorithm of pipelined gossiping.
In particular we found that:
\begin{itemize}
\item The simplest base case, leading to the worst observed performance, 
      is the case which best matches the optimizing algorithm~\ref{algo}. 
      Combining this worst case with the optimization actually leads to a great 
      improvement which, for some values of $N$, raises performance even 
      above the values of the ``best'' base case.
\item Nearly no improvement comes when trying to optimize 
      the ``best'' case.
\end{itemize}

The two above observations seem to suggest that, from
a certain value of $N$ onward, the algorithm of
pipelined gossiping actually \emph{is\/} the ``best'' member of the
family exposed herein. This is suggested for instance from 
Fig.~\ref{compare-efficiency.z+a} and Table~\ref{pots} which show how
even in the best cases, corresponding to a number of processors
equals to a power of 2, there is experimental evidence that,
sooner or later,
the complexity of the problem brings efficiency below 2/3,
the one of pipelined gossiping.
Fig.~\ref{compare-efficiency.z+a} 
shows also that the optimization of the best case in general
does not improve the best case without optimization.
This brought us to the following Conjecture:

Sooner or later, efficiency reaches a value less than or equal to
the one of pipelined gossiping:
\begin{conj}\label{23}
For any $f$ (function transforming parameters like $\mathcal P$),
let us call $\varepsilon_{f,k}$ the efficiency of $f$ in a run with $N=k$
and $\varepsilon_{\textsc{pb}} = 2/3$ the efficiency of the algorithm
of pipelined gossiping;
then there exists an integer $m$ such that 
	$\forall n>m: \varepsilon_{f,n} \leq \varepsilon_{\textsc{pb}}.$
\end{conj}

%

Investigating the above Conjecture will be part of future works.

Of course the optimizing Algorithm~\ref{algo} is not the only one nor the
best possible one.  On the contrary, it is characterized by an
optimization policy which only takes into account the local gain of the
current processor, without any reference to possible global optimization
strategies (e.g., considering also the scenarios that the rest of the
processors are going to face because of the current local choice).
Techniques based on trying alternative solutions and choosing the best
one, possibly considering future consequences of current, local decisions,
may reveal themselves as more appropriate and performant and may be used
to validate the considerations that brought us to Conjecture~\ref{23}.








This paper introduced a family of algorithms depending on a combinatorial parameter and
showed that an optimum exists for its performance is a special case---fully connected and
synchronous systems. Note how such an optimum may exist also in other cases---an open question
that may be dealt with in future work. Should such optima exist, then any tool using
our algorithms could adapt to a change in the communication infrastructure by simply
``loading'' the new optimum. This may have positive relapses on optimal porting of
gossiping services or in mobile systems using gossiping.

\bibliographystyle{latex8}

\end{document}